\documentclass[aps,floatfix,prd,twocolumn,a4paper,10pt,citeautoscript,
preprintnumbers,superscriptaddress,showpacs,longbibliography]{revtex4-1} 

\usepackage[english]{babel} 	
\usepackage{amsmath}        	
\usepackage{amssymb}        	
\usepackage{bm}					
\usepackage{bbm}            	
\usepackage{empheq}
\usepackage{graphicx}       
\usepackage{graphics}       
\DeclareGraphicsExtensions{.jpg,.pdf} 
\usepackage{hyperref}
\hypersetup{colorlinks=true, pdfstartview=FitV, 
linkcolor=blue, citecolor=blue, urlcolor=blue}

\renewcommand{\theequation}{\arabic{equation}}
\newcommand{\Equation}[2]{\begin{equation}\label{#1}#2\end{equation}}
\newcommand{\Align}[2]{\begin{align}\label{#1}#2\end{align}}
\newcommand{\SubAlign}[2]{\begin{subequations}\label{#1}\begin{align}#2\end{align}\end{subequations}}

\newcommand{\bs}{\boldsymbol}
\newcommand{\Figref}[1]{Fig.~\ref{#1}}
\newcommand{\Eqref}[1]{\eqref{#1}}
\newcommand{\groupU}[1]{\mathrm{U}(#1)}  			
\newcommand{\groupZ}[1]{\mathbb{Z}_{#1}} 			
\newcommand{\Exp}[1]{\text{e}^{#1}}
\renewcommand\Re{\mathrm{Re}}
\renewcommand\Im{\mathrm{Im}}
\newcommand{\Grad}{{\bs\nabla}}

\newcommand{\Curl}{{\bs\nabla}\times}

\newcommand{\A}{{\bs A}}
\newcommand{\B}{{\bs B}}
\newcommand{\D}{{\bs \Pi}}

\newcommand{\F}{\mathcal{F}}

\newcommand{\J}{{\bs J}}

\newcommand{\Hc}[1]{\mathrm{H}_{c#1}}

\newcommand{\btheta}{{\bar\theta}}
\newcommand{\bvarphi}{{\bar\varphi}}

\begin{document}
\title{Properties of dirty two-bands superconductors with repulsive 
interband interaction: \texorpdfstring{\\}{} normal modes, length scales, 
vortices and magnetic response}
\author{Julien~Garaud} 
\affiliation{Department of Physics, KTH-Royal Institute of Technology, 
Stockholm, SE-10691 Sweden} 
\affiliation{
Institut Denis-Poisson CNRS/UMR 7013, \\
Universit\'e de Tours - Universit\'e d'Orl\'eans, 
Parc de Grandmont, 37200 Tours, France}
\author{Alberto~Corticelli}  
\affiliation{Department of Physics, KTH-Royal Institute of Technology, 
Stockholm, SE-10691 Sweden} 
\author{Mihail~Silaev}	
\affiliation{Department of Physics and Nanoscience Center, 
University of Jyv\"askyl\"a, P.O. Box 35 (YFL), FI-40014 
University of Jyv\"askyl\"a, Finland}
\author{Egor~Babaev} 
\affiliation{Department of Physics, KTH-Royal Institute of Technology, 
Stockholm, SE-10691 Sweden}

\begin{abstract}
 
Disorder in two-band superconductors with repulsive interband interaction 
induces a frustrated competition between the phase-locking preferences of 
the various potential and kinetic terms. This frustrated interaction can 
result in the formation of an $s+is$ superconducting state, that breaks the 
time-reversal symmetry.
In this paper we study the normal modes and their associated coherence lengths 
in such materials. We especially focus on the consequences of the soft modes 
stemming from the frustration and time-reversal-symmetry breakdown. We find that 
two-bands superconductors with such impurity-induced frustrated interactions 
display a rich spectrum of physical properties that are absent in their clean 
counterparts. 
It features a mixing of Leggett's and Anderson-Higgs modes, and a soft mode 
with diverging coherence length at the impurity-induced second order phase 
transition from $s_{\pm}/s_{++}$ states to the $s+is$ state. Such a soft 
mode generically results in long-range attractive intervortex forces that 
can trigger the formation of vortex clusters. We find that, if such 
clusters are formed, their size and internal flux density have a 
characteristic temperature dependence that could be probed in 
muon-spin-rotation experiments. We also comment on the appearance of 
spontaneous magnetic fields due to spatially varying impurities.

\end{abstract}

\pacs{74.25.Dw,74.20.Mn,74.62.En}
\date{\today}
\maketitle

\section{Introduction}

The discovery of iron-based superconductors motivated research on two-band 
superconductors where the pairing between electrons is produced by interband 
electron-electron repulsion \cite{Mazin.Singh.ea:08,Chubukov.Efremov.ea:08,
Hirschfeld.Korshunov.ea:11}. 
Such systems tend to form a state with two $s$-wave gaps 
$\Delta_i=|\Delta_i|\Exp{i\theta_i}$ (with $i=1,2$), for which the relative 
phase differs by $\pi$ (that is, $\theta_2=\theta_1+\pi$). This superconducting 
state with a sign change between the gap functions is called $s_\pm$, in 
contrast to the more commonly studied $s_{++}$ state, which has a zero relative 
phase ($\theta_1=\theta_2$). 
The $s_\pm$ superconducting state behaves non trivially when disorder
is added. It is indeed known that, under certain conditions, impurities 
induce a crossover from the $s_{\pm}$ to the $s_{++}$ state. At temperatures 
sufficiently close to the critical temperature $T_c$, the transition from 
the $s_\pm$ to the $s_{++}$ state is realized as a direct crossover, with 
little thermodynamic features, where one of the gap function is completely 
suppressed \cite{Efremov.Korshunov.ea:11}. It was nonetheless recently 
demonstrated that, due to competing kinetic and potential terms, inhomogeneous 
states such as vortices or screening currents become structurally nontrivial 
in the vicinity of that crossover \cite{Garaud.Silaev.ea:17a,
Garaud.Corticelli.ea:18}.

At lower temperatures, the impurity-induced transition to the $s_{++}$ state 
occurs via an intermediate state where the inter-component relative phase 
is different from $0$ and $\pi$ \cite{Bobkov.Bobkova:11,Stanev.Koshelev:14}. 
This state is called $s+is$ state. It spontaneously breaks the time-reversal 
symmetry, and is separated from the standard $s_\pm/s_{++}$ states by a 
second-order phase transition (at mean-field level). However quantitative 
calculations of the phase diagram demonstrated \cite{Silaev.Garaud.ea:17} 
that the impurity induced $s+is$ state occupies a vanishingly small region of 
the phase diagram and is unlikely to be observable directly. Note that this 
statement applies only to weak-coupling mean-field theory of dirty two-band 
system. This behavior is drastically different from that found in systems with 
three or more interacting bands, where the $s+is$ state appears as a result 
of the frustrated interband repulsive pairing \cite{Ng.Nagaosa:09,Stanev.Tesanovic:10,
Carlstrom.Garaud.ea:11a,Maiti.Chubukov:13,Boeker.Volkov.ea:17}.

In this work we demonstrate that, even if the $s+is$ state occupies a very small 
region, its mere presence on the phase diagram can still have important 
consequences. Indeed, as previously stated the $s+is$ state spontaneously breaks 
the time-reversal symmetry. Thus, in addition to the usual $\groupU{1}$ symmetry, 
the $s+is$ state also breaks the discrete $\groupZ{2}$ symmetry associated with 
the time-reversal operations. In other words, since the relative phase between 
the gaps is neither $0$ nor $\pi$, complex conjugation leads to another state 
that cannot be rotated back to the initial state by $\groupU{1}$ transformation.
Because $s_\pm/s_{++}$ are different on symmetry grounds from the $s+is$ 
state, at the mean-field level the phase transition is second order. This 
implies that there is a divergent coherence length inside the superconducting 
state on both sides of the $s+is$ domain \cite{Silaev.Garaud.ea:17}.

Here we demonstrate that the emerging soft normal mode with divergent coherence 
length is not only associated with the relative phase (Leggett's mode), but also 
with the amplitude (Higgs modes). This leads to the situation that $s_{\pm}/s_{++}$ 
states adjacent to the $s+is$ domain should acquire unconventional properties 
associated with the static and dynamic fluctuations, the nature of topological 
excitations, and the magnetic response to an external applied field. Therefore 
dirty two-band superconductors with a repulsive interband interaction have a much 
more complex behavior than the well studied standard $s_{++}$ state in clean systems 
where, by contrast, the existence of soft modes away from superconducting phase 
transition, requires only weak interband coupling \cite{Silaev.Babaev:11}.

The paper is organized as follows. In section \ref{Sec:GL}, starting from the 
microscopic Usadel theory of dirty two-band superconductors, we provide a 
detailed derivation of the corresponding two-band Ginzburg-Landau model, and 
discuss the essential properties of the phase diagram. 
Next, section \ref{Sec:Perturbation} is devoted to the complete analysis of 
the linearized theory. This provides the framework to describe the behavior 
of the coherence lengths and their associated normal modes, across the 
different phases of the phase diagram. The derived perturbation operator can 
also be used to determine the upper critical fields of such dirty two-band 
superconductors. This is discussed separately in an appendix.
The perturbation operator features a divergent length scale in the vicinity 
of the second order phase transition to the $s+is$ phase. The existence 
of such a soft mode can result in long-range attractive intervortex forces. 
In section \ref{Sec:Vortices} we investigate this property beyond the linear 
regime regime, and demonstrate that vortex clusters can form in the vicinity 
of the $s+is$ phase, and that they feature specific temperature dependent 
properties.

Readers who are not interested in technical details of the analysis of normal 
models but are interested in properties of vortex states and their possible 
experimental manifestations can, after Sec.~\ref{Sec:GL}, directly proceed to 
Sec.~\ref{Sec:Vortices}.

\section{Ginzburg-Landau model derived from the Usadel equations}
\label{Sec:GL}

We investigate the properties of the superconducting states, their 
characteristic length scales and vortex structures within a weak-coupling 
model of two-band superconductors with high concentration of impurities. 
Such material can be described by a system of two Usadel equations coupled 
together by interband impurity scattering terms \cite{Gurevich:03}:
\Align{Eq:Usadel}{
\omega_n f_i &= \frac{D_i}{2} \big(g_i\D^2 f_i - f_i \nabla^2 g_i\big) 	
			 +  \Delta_i g_i									\nonumber \\
 			 &~~~+\sum_{j\neq i}\gamma_{ij} ( g_if_j - g_jf_i) 	\,,
}
where $\omega_n = (2n+1)\pi T $, with $n\in\mathbb{Z}$, are the fermionic 
Matsubara frequencies. $T$ stands for the temperature, $D_i$ are the 
electron diffusivities, and $\gamma_{ij}$ are the interband scattering rates. 

The quasi-classical propagators $f_i$ and $g_i$, which are, respectively, the 
anomalous and normal Green's functions in each band, obey the normalization 
condition $|f_i|^2 + g_i^2 =1$. The two components 
$\Delta_j=|\Delta_j|e^{i\theta_j}$ of the order parameter are determined by 
the self-consistency equations
\Equation{Eq:SelfConsistency}{
 \Delta_i =2\pi T  \sum_{n=0}^{N_d} 
 \sum_{j} \lambda_{ij} f_{j} (\omega_n),
}
for the Green's functions that satisfy the Usadel equation Eq.~\Eqref{Eq:Usadel}. 
Here, $N_d=\Omega_d/(2\pi T)$ is the summation cut-off at the Debye frequency 
$\Omega_d$. In the self-consistency equation \Eqref{Eq:SelfConsistency}, the 
diagonal elements $\lambda_{ii}$ of the coupling matrix $\hat{\lambda}$ describe 
the intraband pairing, while the interband interaction is determined by the 
off-diagonal terms $\lambda_{ij}$ ($j\neq i$). The interband coupling parameters 
and impurity scattering amplitudes satisfy the symmetry relation \cite{Gurevich:03}:
\Equation{Eq:Symmetry}{
 \lambda_{ij}= - \lambda_J/n_i~~\text{and}~~\gamma_{ij}=\Gamma n_j \,,
} 
where $\lambda_J$ and $\Gamma>0$. The impurity scattering rate is given in units 
of $T_c$, $n_i=N_i/(N_1+N_2)$ are the relative densities of states, and $N_{1,2}$ 
are the partial densities of states in the two-bands.

In general, the $s_\pm$ state is not favored by the impurity scattering, 
which tends to average out the order parameter over the whole Fermi surface, 
suppressing the critical temperature. Still, provided the interband pairing 
interaction is weak, superconductivity can be transformed into a $s_{++}$ state 
and survive even in the limit $\Gamma\gg T_{c0}$, characterized by the critical 
temperature $T_{c\infty}$ which reads as \cite{Stanev.Koshelev:14,
Hirschfeld.Korshunov.ea:11}: 
\Equation{Eq:Tc}{
 \ln (T_{c0}/T_{c\infty}) =  N_1 (w_{11}+w_{12}) + N_2 (w_{22}+w_{21}) \,,
}
where $T_{c0}$ is the critical temperature in the absence of interband scattering,
$\hat w = \hat \Lambda^{-1}- z^{-1}\hat I $ and $z$ is the maximal eigenvalue 
of the coupling matrix $\hat\Lambda$ with the elements $\lambda_{kk^\prime}$. 
In order to avoid a drastic suppression of the critical temperature in the 
$s_{++}$ state, according to Eq.~\Eqref{Eq:Tc}, the interband interaction 
$\lambda_J$ should be sufficiently weak. To derive a criterion, note that 
$N_1 w_{11} + N_2 w_{22} > 0$, so that the r.h.s. of the Eq.~\Eqref{Eq:Tc} 
is larger than $N_1 w_{12} + N_2 w_{21} = \lambda_J /(\lambda_{11}\lambda_{22})$. 
Therefore, in order to have a $T_{c\infty}$ which is not much smaller than 
$T_{c0}$, we require the following condition 
$\lambda_J /(\lambda_{11}\lambda_{22}) <1 $ to be satisfied.

\subsection{Ginzburg-Landau expansion}

The two-band Ginzburg-Landau (GL) expansion is an expansion in two small gaps 
and small gradients [not to be confused with a single-parameter expansion 
$\tau=(1-T/T_c)$]. A detailed discussion of the formal validity of multiband 
expansions in the context of clean system can be found in \cite{Silaev.Babaev:12}. 
It was demonstrated in Ref.~\onlinecite{Silaev.Garaud.ea:17} that for dirty 
systems, in the region of its applicability, the Ginzburg-Landau model gives 
a phase diagram that matches that of the microscopic Usadel theory.
Here, we provide the full derivation of the GL expansion including gradient terms.
In the case of a dirty system, by inverting the self-consistency equation 
\Eqref{Eq:SelfConsistency}, it is found that: 
\Equation{Eq:SelfConsistency:Inv}{
 2\pi T \sum_{n=0}^{N_d} f_{i}(\omega_n) = 
 \frac{ \lambda_{jj}\Delta_k - \lambda_{ij}\Delta_j}{\mathrm{det}\hat{\lambda}} 
~~~~\text{and}~j\neq i\,.
}
and defining the expansion for the $f_i$ from the Usadel equation 
\Eqref{Eq:Usadel}. In the first approximation we put $g^{(0)}_i=1$ 
(at $\omega_n>0$) and thus find 
\Equation{Eq:fZeroOrder}{
f^{(1)}_i=\frac{ \gamma_{ij}\Delta_j + ( \omega_n + \gamma_{ji} )\Delta_i }
{\omega_n ( \omega_n+ \gamma_{ij}+\gamma_{ji})}
~~~~\text{and}~j\neq i\,.
}
The corrections $f^{(3)}_i$ from the non-linear terms in Eq.~\Eqref{Eq:Usadel} 
are found by neglecting the gradients from which follows the general relation
\Equation{Eq:fExact}{
  f_i=\frac{\Delta_i (\omega_n + \gamma_{ji}g_i) + \gamma_{ij} \Delta_j g_j }
  { \omega_n (\omega_n + \gamma_{ij}g_j + \gamma_{ji}g_i  ) } g_i   \,.
}
Then, when taking into account the corrections, $g_i =1 - |f_i^{(1)}|^2/2$, 
and this yields
\Align{Eq:fNonlinearCorr}{
f^{(3)}_i=&-\frac{ |f^{(1)}_i|^2\Delta_i\left[\left(\omega_n +\gamma_{ji}\right)^2 
  			+ \gamma_{ij} (\omega_n + 2\gamma_{ji}) \right] }
   {2\omega_n (\omega_n+\gamma_{ij}+\gamma_{ji})^2} \nonumber \\ 
  & - \frac{|f^{(1)}_i|^2 \Delta_j(\omega_n+\gamma_{ij})\gamma_{ij}}
   {2\omega_n (\omega_n+\gamma_{ij}+\gamma_{ji})^2}  \nonumber\\ 
  & + \frac{|f^{(1)}_j|^2 \gamma_{ij}(\omega_n+\gamma_{ji}) (\Delta_i-\Delta_j)  }
   {2\omega_n (\omega_n+\gamma_{ij}+\gamma_{ji})^2}		\,.
}
Finally, combining the equations \Eqref{Eq:fExact} and \Eqref{Eq:fNonlinearCorr} 
yields the non-linear terms in the Ginzburg-Landau expansion. The corrections 
$f^{(g)}_i$ from the gradient terms are obtained by linearizing Usadel equation 
\Eqref{Eq:Usadel}, with respect to the corrections $f^{(g)}_i$. This yields 
\Align{Eq:fGradients}{
  f^{(g)}_i =& \frac{D_i (\omega_n +\gamma_{ji})^2 + D_j\gamma_{ij}\gamma_{ji}}
  {2\omega_n^2(\omega_n+\gamma_{ij}+\gamma_{ji})^2} {\bm \Pi}^2\Delta_i
  \nonumber  \\
  &+\frac{ \gamma_{ij}
  \left[ D_i (\omega_n +\gamma_{ji} )+ D_j ( \omega_n+ \gamma_{ij})\right] }
  {2\omega_n^2(\omega_n+\gamma_{ij}+\gamma_{ji})^2} {\bm \Pi}^2\Delta_j \,.
}
Finally, the Ginzburg-Landau functional reads as
\SubAlign{Eq:FreeEnergy}{
\frac{\F}{\F_0} =&
\sum_{j=1}^2\Big\{
 	\frac{k_{jj}}{2}\left|\D\Delta_j \right|^2
 	+a_{jj}|\Delta_j|^2+\frac{b_{jj}}{2}|\Delta_j|^4\Big\}  
 	\label{Eq:FreeEnergy:Self}	\\
   &+\frac{k_{12}}{2}
 	\Big((\D\Delta_1)^*\D\Delta_2+(\D\Delta_2)^*\D\Delta_1 \Big)
	\label{Eq:FreeEnergy:Mixed}	\\
   &+2\left(a_{12}+c_{11}|\Delta_1|^2+c_{22}|\Delta_2|^2\right)
   \Re\big(\Delta_1^*\Delta_2\big)
	\label{Eq:FreeEnergy:Interaction1}	\\
   &+\left(b_{12}+c_{12}\cos2\theta_{12}\right)|\Delta_1|^2|\Delta_2|^2    
   +\frac{\B^2}{2}
	\label{Eq:FreeEnergy:Interaction2}	\,.
}
Here, $\theta_{12}=\theta_2-\theta_1$ stands for the relative phase between
the complex fields $\Delta_j=|\Delta_j| e^{i\theta_j}$ that represent the 
superconducting gaps in the different bands. The two gaps in the different 
bands are electromagnetically coupled by the vector potential $\A$ of the 
magnetic field $\B=\bs\nabla\times\A$, through the gauge derivative 
$\D\equiv\Grad+iq\A$. 
The coefficients of the Ginzburg-Landau functional $a_{ij}$, $b_{ij}$, 
$c_{ij}$ and $k_{ij}$ can be calculated from a given set of input microscopic 
parameters $\lambda_{ij}$, $D_i$, $T$ and $\Gamma$ of the microscopic 
self-consistency equation. Their explicit formulas are listed in 
Appendix~\ref{App:Coefficients}.

As can be seen in Eq.~\Eqref{Eq:fGradients}, the coefficients of the 
gradient terms depend on both electronic diffusivity coefficients $D_1$ 
and $D_2$. Clearly, the parameter space can be reduced by absorbing one 
of the electronic diffusivity coefficient in the gradient term. Without 
any loss of generality, we choose $D_1$ to be the largest diffusivity 
coefficient ($D_1>D_2$). Thus, in the dimensionless units, the coefficients 
of the gradient term depend only on the ratio of diffusivities, or 
\emph{relative diffusion constant} $r_d=D_2/D_1<1$.
The free energy \Eqref{Eq:FreeEnergy} is expressed in terms of dimensionless 
quantities, so the coupling constant $q$ should not be confused with $ 2\pi/\Phi_0$.  
In such units, the coupling constant $q$ instead parametrizes the penetration 
depth of the magnetic field (see units detail below). The dimensionless units 
are defined as:
\Equation{Eq:Dimensionless}{
\Grad= \xi_0\tilde{\Grad}\,,~~ 
\A=\tilde{\A}/ B_0\xi_0\,,~~ 
\Delta=\tilde{\Delta}/ T_{c}\,,
}
where the variables with tilde are the dimensionful quantities. Therefore
$\xi_0 = \sqrt{D_1/T_{c}}$ is the new unit of the length, and 
$B_0 = T_c \sqrt{4\pi N_1}$ the unit of the magnetic field (here $N_1$ is the 
density of states in the first band). The free energy is then scaled by 
${\F_0} =B_0^2/4\pi $, while the electromagnetic coupling constant becomes 
$q = 2\pi B_0 \xi_0^2/\Phi_0$. 

In these new units, the London penetration 
length $\lambda_L$ is given by 
$\lambda^{-2}_L=q^2(k_{ii}\Delta_{i0}^2+2 k_{12}\Delta_{10}\Delta_{20})$,
where $\Delta_{i0}$ is the bulk value of the dimensionless gap. 
Correspondingly the gauge field coupling constant is
$q=\lambda_L(k_{ii}\Delta_{i0}^2+2 k_{12}\Delta_{10}\Delta_{20})^{-1/2}$.
%
Eventually, for a given set of input microscopic parameters, $\lambda_{ij}$, 
$\Gamma$, $r_d$, and $T$ close to $T_c$, we can reconstruct the coefficients 
and investigate the ground-state properties of the GL theory by minimizing the 
free energy \Eqref{Eq:FreeEnergy} with respect to $|\Delta_j|$ and $\theta_{12}$.

\subsection{Phase diagrams}

The mean-field phase diagram of the dirty two-band superconductors was 
calculated in \cite{Silaev.Garaud.ea:17}, using both Usadel and Ginzburg-Landau 
formalisms. It was demonstrated there, that the phase diagrams are quantitatively 
similar within the range of validity of GL expansion. Here we briefly outline the 
structure of the diagram within the Ginzburg-Landau model.  Knowing the coefficients 
(see details in Appendix \ref{App:Coefficients}) of the microscopically derived 
Ginzburg-Landau functional \Eqref{Eq:FreeEnergy}, allows to investigate the 
ground-state properties of dirty two-bands superconductors by minimizing the 
free energy \Eqref{Eq:FreeEnergy} with respect to $|\Delta_j|$ and $\theta_{12}$.

\begin{figure}[!htb]
\hbox to \linewidth{ \hss
\includegraphics[width=0.99\linewidth]{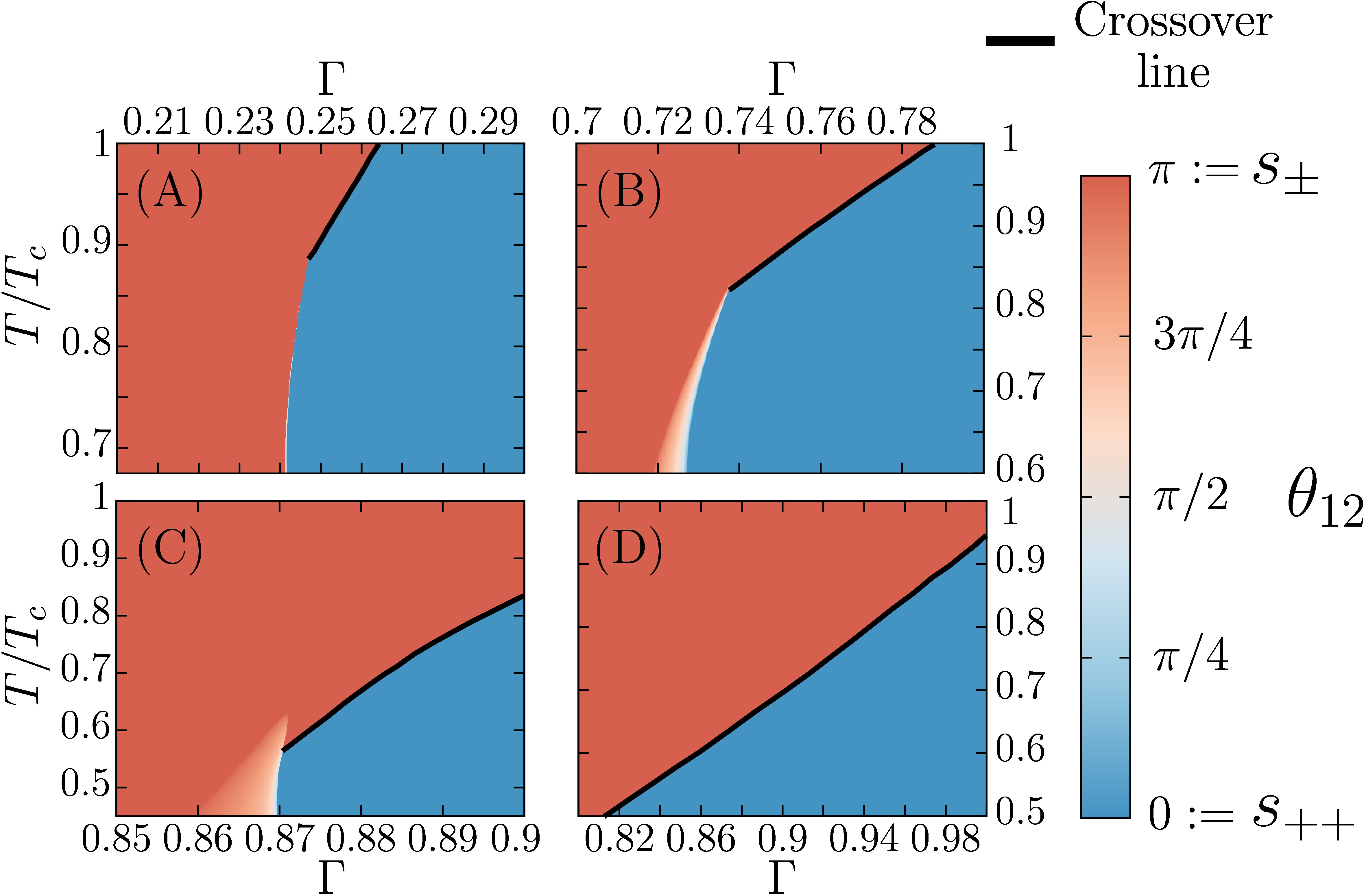}
\hss}
\caption{
Phase diagrams of the Ginzburg-Landau free energy \Eqref{Eq:FreeEnergy} 
describing two-band superconductors with interband impurity scattering. 
These show the values of the lowest-energy state relative phase 
$\theta_{12}=\theta_2-\theta_1$ between the components of the order 
parameter, as function of temperature and interband scattering $\Gamma$. 
The different panels correspond to different values of the coupling 
matrix $\hat{\lambda}$. Panels (A), (B), and (C) respectivelly correspond 
to nearly degenerate bands with $\lambda_{11}=0.29$ and $\lambda_{22}=0.3$ 
with weak $\lambda_{12}=\lambda_{21}=-0.01$, intermediate 
$\lambda_{12}=\lambda_{21}=-0.05$ and strong $\lambda_{12}=\lambda_{21}=-0.1$
repulsive interband pairing interaction. The last panel (D) describes the 
case of intermediate band disparity $\lambda_{11}=0.25$ and $\lambda_{22}=0.3$ 
with intermediate $\lambda_{12}=\lambda_{21}=-0.05$ repulsive interband 
pairing interaction. The solid black line shows the zero of $\Delta_2$, 
that is the crossover between $s_\pm$ and $s_{++}$ states. 
In panels (A), (B) and (C), the crossover line is attached to a dome of 
time-reversal symmetry breaking $s+is$ state. In the panel (D), the crossover 
line does not connect to an $s+is$ state. 
}
\label{Fig:Diagram}
\end{figure}

The phase diagrams are constructed in the plane of parameters $\Gamma,T$ of a 
two-band superconductor with interband impurity scattering. For that 
purpose, we numerically minimize the free energy \Eqref{Eq:FreeEnergy} 
using a nonlinear conjugate gradient algorithm. The results displayed in 
\Figref{Fig:Diagram} demonstrate the role of impurities on the ground 
state properties, for various representative cases. Namely nearly degenerate 
bands with weak (A), intermediate (B) and strong (C) repulsive interband 
pairing interactions (as compared to the intraband couplings). Also, 
we consider the case of intermediate band disparity with intermediate
interband coupling (D).
 
Those diagrams illustrate the now well understood fact that, in two-band 
superconductors, disorder may induce a transition from the $s_{\pm}$ state 
(the red regions with $\theta_{12}=\pi$ in \Figref{Fig:Diagram}) to the $s_{++}$ 
state (the blue regions with $\theta_{12}=0$ in \Figref{Fig:Diagram}) 
\cite{Efremov.Korshunov.ea:11,Bobkov.Bobkova:11,Stanev.Koshelev:14,
Silaev.Garaud.ea:17}. The transition can occur in two qualitatively different 
ways. Either via a direct crossover (denoted by a solid black line) when one 
of the superconducting gap vanishes as a function of impurity concentration 
\cite{Efremov.Korshunov.ea:11}, or via the intermediate complex $s+is$ state 
that break time-reversal symmetry with $\theta_{12} \ne 0, \pi$. The crossover 
occurs without additional symmetry breaking while the transition via an $s+is$ 
state spontaneously breaks the time-reversal symmetry, and both $s_{\pm}/s+is$ 
and $s_{++}/s+is$ transitions lines are of the second order, at the mean-field 
level \cite{Silaev.Garaud.ea:17}. As was mentioned in the Introduction, 
the existence of the second-order phase transition on the phase diagram dictates 
that there is softening of one of the normal modes near that transition. 
This softening has a number of possible physical consequences. That motivates 
the study performed in the next section, where we consider the normal modes of 
this system.

\section{Linear analysis: normal modes and coherence lengths} 
\label{Sec:Perturbation}

An analysis of the perturbation operator around classical solutions 
such as the ground state, or the normal state provides important informations
such as the length scales of the theory, the zero modes or the upper critical 
field. To facilitate this analysis, we find it convenient here to rewrite the 
Ginzburg-Landau free energy \Eqref{Eq:FreeEnergy} in terms of a new \emph{rotated} 
field basis (linear combination) that eliminates the mixed gradient terms.

\subsection{Elimination of mixed gradient terms} 

Because it features mixed gradient terms, the original basis for the 
superconducting degrees of freedom is quite inconvenient to work with.
This is why it is worth rewriting the model using a linear combination of 
the components of the order parameter that diagonalizes the kinetic terms:
\Equation{Eq:NewFields}{
\psi_1=\sqrt{k_{11}}\Delta_1+\sqrt{k_{22}}\Delta_2\,,~~
\psi_2=\sqrt{k_{11}}\Delta_1-\sqrt{k_{22}}\Delta_2\,.
}
Within this new basis, we refer to as \emph{rotated basis}, the kinetic term 
has a much simpler form. The potential, on the other hand becomes more involved. 
Yet it is a convenient basis to deal with, for the determination of the physical 
length scales as well as describing various unusual properties. In the new 
rotated field basis, the free energy reads as 
\SubAlign{Eq:FreeEnergy:Rotated}{
&\F=\sum_{j=1}^2\Big\{
 	\frac{\kappa_{j}}{2}\left|\D\psi_j \right|^2
 	+\alpha_{jj}|\psi_j|^2+\frac{\beta_{jj}}{2}|\psi_j|^4\Big\}  
 	\label{Eq:FreeEnergy:Rotated:Self}	\\
   	&+2\left(\alpha_{12}+\gamma_{11}|\psi_1|^2|+\gamma_{22}\psi_2|^2\right)
   	|\psi_1||\psi_2|\cos\varphi_{12}
   	\label{Eq:FreeEnergy:Rotated:Interaction1}	\\
	&+\left(\beta_{12}+\gamma_{12}\cos2\varphi_{12}\right)|\psi_1|^2|\psi_2|^2
	+\frac{\B^2}{2}
	\label{Eq:FreeEnergy:Rotated:Interaction2}	\,,
}
with the \emph{rotated} superconducting degrees of freedom 
$\psi_j=|\psi_j|\Exp{i\varphi_j}$, and $\varphi_{12}=\varphi_2-\varphi_1$, 
and the coefficients for the kinetic term are now
\Equation{Eq:CoeffKinetic}{
\kappa_1=\frac{\sqrt{k_{11}k_{22}}+ k_{12}}{2\sqrt{k_{11}k_{22}}}
~~~~\text{and}~~~~
\kappa_2=\frac{\sqrt{k_{11}k_{22}}- k_{12}}{2\sqrt{k_{11}k_{22}}}
\,.
}
Upon some algebraic manipulations, all coefficients $\alpha_{ij}$, $\beta_{ij}$, 
$\gamma_{ij}$ of the potential, are expressed in terms of the coefficients 
$a_{ij}$, $b_{ij}$, $c_{ij}$ and $k_{ij}$ of the original Ginzburg-Landau 
functional \Eqref{Eq:FreeEnergy}. Detailed expressions of new parameters 
can be found in Appendix~\ref{App:CoefficientsRewritten}. Within the framework 
of new \emph{rotated} variables \Eqref{Eq:NewFields}, the Ginzburg-Landau 
equations have no mixed-gradients and read as 
\Equation{Eq:EOM:Rotated}{
\D^2\psi_j
=2\frac{\partial V}{\partial\psi_j^*}\,.
} 
The variation of the free energy \Eqref{Eq:FreeEnergy:Rotated} with respect 
to the vector potential $\A$, determines Amp\`ere's equation $\Curl\B+\J=0$. 
There, the total current is the superposition of the partial currents 
($\J=\sum_i\J^{(i)}$) that reads as 
\Equation{Eq:Currents:Rotated}{
\J^{(i)} = q\kappa_i\Im\big(\psi_i^*\D\psi_i\big)\,.
}

The reparametrization \Eqref{Eq:NewFields} simplifies drastically the 
Ginzburg-Landau equations as there is no more coupling of the components 
through mixed gradients. However, this comes with the price of more 
complicated potential terms. This is actually a minor issue, since the 
ground state within the rotated basis, can easily be determined from the 
one in the original field basis according to the formulas
\Align{Eq:Transform}{
|\psi_1|^2&=k_{11}|\Delta_1|^2+k_{22}|\Delta_2|^2
	+2\sqrt{k_{11}k_{22}}|\Delta_1||\Delta_2|\cos\theta_{12}	
	\,, \nonumber\\
|\psi_2|^2&=k_{11}|\Delta_1|^2+k_{22}|\Delta_2|^2
	-2\sqrt{k_{11}k_{22}}|\Delta_1||\Delta_2|\cos\theta_{12}	
	\,, \nonumber\\
\varphi_{12}&=\tan^{-1}\left(
\frac{-2\sqrt{k_{11}k_{22}}|\Delta_1||\Delta_2|\sin\theta_{12}}
	 {k_{11}|\Delta_1|^2-k_{22}|\Delta_2|^2}	\right)\,.
}
 
To understand the role of excitations, as well as the fundamental length 
scales of the Ginzburg-Landau free energy \Eqref{Eq:FreeEnergy:Rotated}, 
it can be rewritten in terms of gauge invariant quantities (i.e. in terms 
of \emph{charged} and \emph{neutral} modes (see a general discussion in 
the context of a simpler model in \cite{babaev2002hidden,babaev2002vortices})
by expanding the kinetic term in \Eqref{Eq:FreeEnergy:Rotated:Self} and 
using \Eqref{Eq:Currents:Rotated}:
\Align{Eq:GLRewritten}{
   &\F= \frac{1}{2}(\Curl \A)^2 + \frac{\J^2}{2q^2\varrho^2} 
   +\sum_{a}\frac{\kappa_a}{2}(\Grad|\psi_a|)^2
    \nonumber\\
	&+\frac{\kappa_1\kappa_2|\psi_1|^2|\psi_2|^2}{2\varrho^2}
	(\Grad\varphi_{12})^2 
	+V(|\psi_1|,|\psi_2|,\varphi_{12})
	\,.
}
Here again $\varphi_{12}=\varphi_2-\varphi_1$ stands for the relative 
phase between the condensates, and $\varrho^2=\sum_{i}\kappa_i|\psi_i|^2$. 
For this rewriting, we used the supercurrent defined from the Amp\`ere's 
equation $\Curl\B+\J=0$, that reads
\Equation{Eq:Currents2}{
  \J/q= q\varrho^2 \A+\sum_{i}\kappa_i|\psi_i|^2
  \Grad\varphi_a  \,.
}
As discussed below, due to absence of mixed gradient terms, this formulation 
allows an easier calculation of the length scales and a better interpretation 
of the corresponding normal modes.

\subsection{Coherence lengths and perturbation operator} 
 
The length scales that characterize matter fields are called coherence 
lengths. Fundamentally the coherence length $\xi$ associated to a field 
$\Psi({r})$ is defined through the exponent that characterizes how, 
from a small perturbation, the field recovers its ground state value 
$\bar{\Psi}$ (see, e.g., \cite{tinkham,plischke,svistunov}). That is, 
from a perturbation $\Psi ({r}) \approx \bar{\Psi}$, the field recovers 
according to the asymptotic behavior 
\Equation{coh}{
 \Psi ({ r})-\bar{\Psi}\propto e^{-\frac{r}{\xi}} \ \ \ \  (r\gg\xi)\,.
}
Note that, typically in the context of superconductivity the definition 
of the coherence length has an extra $\sqrt{2}$ factor \cite{tinkham}, 
while this factor is absorbed into the definition of $\xi$ in other contexts. 
Here we follow the more general definition and absorb this factor into $\xi$. 
Note also that for the simplest Ginzburg-Landau model the coherence length 
is occasionally indirectly assessed, for example through overall vortex core 
size or from the slope of the order parameter near the center of the vortex 
core. Only in some special cases all these estimates give consistent results.  
For example, even in the simplest single-component $s$-wave superconductors, 
away from $T_c$ all these definitions give inconsistent results
\cite{Gygi.Schlueter:91}. 
In a multi-component systems, the length scales physics is more complicated 
so they should not be a priori expected to be easily assessable from such 
quantities as the order parameter slope near the origin. 
Another consequence of intercomponent interactions, is that it cannot be 
expected that independent coherence lengths are associated with single fields 
$\Delta_{j}$. Instead one can expect to find linear combinations of the complex 
fields that recover from a perturbation with different exponential laws 
\Eqref{coh} and therefore are characterised by different coherence lengths. 
In general, in multi-component GL models, determination of the various coherence 
lengths cannot be done analytically, except in the cases of weak interband 
interaction, where the intercomponent interactions can be addressed perturbatively 
\cite{Carlstrom.Babaev.ea:11}. Thus generic determination of the coherence 
lengths has to be carried out numerically. 

To determine the coherence lengths one thus consider the small perturbations 
in all relevant field degrees of freedom around a physical solution, and 
linearize the theory around that solution. Such a physical solution is for 
example the ground-state, the normal state, etc. The eigenvalue spectrum of 
the infinitesimal perturbation operator are the (squared) masses of the normal 
modes, and the coherence lengths are defined as the inverse masses. Thus, the 
eigenspectrum of the obtained (linear) differential operator determines the 
masses of the normal modes and consequently their corresponding length scales.
In the single-component limit that corresponds to the standard calculation  \cite{tinkham}. 
By contrast, the model we consider here has four degrees of freedom associated 
with the matter fields: two moduli and two phases of the complex fields.

If one neglect the coupling to vector potential then the sum of the phases 
forms a mode with zero mass (the Goldstone mode), since it is associated 
with a broken $\groupU{1}$ symmetry. When coupling to the vector potential 
is included this mode becomes massive via the London-Anderson-Higgs mechanism. 
The inverse of that mass is the London's magnetic field penetration length. 
For the simplest two-band $s_{++}$ material the phase difference constitute 
another massive mode that, in a dynamical context, is called the Leggett's 
mode \cite{leggett}. In a static case the length scale associated with this 
mode (i.e. the length scale at which the phase difference recovers from a 
perturbation) is also called Josephson length. However it was discussed in 
clean three-band superconductors, that when time-reversal symmetry is broken, 
there is no Leggett-type (phase-only) mode, and instead the phase difference 
mode is hybridized (i.e. mixed) with the density (Higgs) modes 
\cite{Carlstrom.Garaud.ea:11a,Maiti.Chubukov:13,Stanev:12,Marciani.Fanfarillo.ea:13}. 
Below, we find that in impurities-induced $s+is$ case the modes are mixed as 
well. As dictated by the theory of the mean-field second order phase transitions, 
mass of one of the modes should go to zero at the superconducting phase transition 
(indeed at this transition $\groupZ{2}$ symmetry is broken and thus there is 
divergence of one of the coherence lengths, while other length scales should 
remain finite). In Ref.~\onlinecite{Silaev.Garaud.ea:17} it was demonstrated 
that the transition to the $s+is$ state from $s_{++}$ or $s_\pm$ state is second 
order at the mean-field level. This dictates that there should be a divergent
coherence length at that transition as well.

The perturbation theory is constructed as follows. The fields are expanded 
in series of a small parameter $\epsilon$: $\psi_i=\sum_a\epsilon^a\psi_i^{(a)}$ 
and collected order by order in the functional. The zeroth order is the 
original functional, while the first order is identically zero provided the 
leading order in the series expansion satisfies the equations of motion. 
Because we expand near a classical state (for example the ground state), 
physically relevant correction thus appear at the order $\epsilon^2$ of 
the expanded Ginzburg-Landau functional.
The length scale analysis is done by applying the previously discussed 
perturbative theory to the case where the leading order is the ground-state.

We choose the following expansion in small perturbations around the ground 
state
\Equation{Eq:expansion}{
|\psi_i|=u_i+\frac{\epsilon f_i}{\sqrt{\kappa_i}}\,,~~
\varphi_{12}=\bvarphi+\epsilon\sqrt{\frac{\kappa_1u_1^2+\kappa_2u_2^2}
{\kappa_1\kappa_2u_1^2u_2^2}}\phi\,.
}
where $u_i$ and $\bvarphi$ denote the ground state while $f_i$ and $\phi$, 
stand for the perturbations. $\epsilon$ is the arbitrarily small parameter 
of the series expansion. Collecting the perturbations in a single vector 
$\Upsilon=(f_1,f_2,\phi)^T$, the term which is second order in $\epsilon$ 
in the Ginzburg-Landau functional \Eqref{Eq:GLRewritten} reads as:
\Equation{Eq:perturbation:operator}{
\frac{1}{2}\Upsilon^T\left(\Grad^2+{\cal M}^2  \right)\Upsilon\,,
} 
where the diagonal entries of the (squared) mass matrix are: 
\SubAlign{Eq:perturbation:1}{
{\cal M}^2_{f_if_i}&=\frac{2}{\kappa_i}\Big(
	\alpha_{ii}+3\beta_{ii} u_i^2+(\beta_{12}+\gamma_{12}\cos2\bvarphi) u_j^2
\nonumber \\
&~~~~~+6\gamma_{ii} u_1u_2\cos\bvarphi \Big)\\
{\cal M}^2_{\phi\phi}&=
-\frac{\kappa_1u_1^2+\kappa_2u_2^2}{\kappa_1\kappa_2u_1^2u_2^2}
\Big( 4\gamma_{12} u_1^2u_2^2\cos2\bvarphi		
\nonumber \\
&~~~~~+2(\alpha_{12}+\gamma_{11}u_1^2+\gamma_{22}u_2^2)u_1u_2\cos\bvarphi\Big)	\,,
}
and the off-diagonal elements are
\SubAlign{Eq:perturbation:2}{
{\cal M}^2_{f_1f_2}&=\frac{1}{\sqrt{\kappa_1\kappa_2}}
	\Big(4(\beta_{12}+\gamma_{12}\cos2\bvarphi)u_1u_2
\nonumber \\
&~~~~~+2(\alpha_{12}+3\gamma_{11}u_1^2+3\gamma_{22}u_2^2)\cos\bvarphi \Big)\\
{\cal M}^2_{f_i\phi}&=
-\sqrt{\frac{\kappa_1u_1^2+\kappa_2u_2^2}{\kappa_1^2\kappa_2u_1^2u_2^2}}
\Big( 4\gamma_{12} u_iu_j^2\sin2\bvarphi		
\nonumber \\
&~~~~~+2(\alpha_{12}+3\gamma_{ii}u_i^2+\gamma_{jj}u_j^2)u_j\sin\bvarphi	\Big)	\,,
}
with $j\neq i$. From here, the benefit of using the rotated basis for the 
fields \Eqref{Eq:NewFields} together with the gauge invariant formulation 
\Eqref{Eq:GLRewritten}, becomes rather clear. Indeed, within that formulation, 
the perturbation operator \Eqref{Eq:perturbation:operator} has off-diagonal 
terms coupling various excitations only in the mass matrix.
It is worth emphasizing here, that the perturbation operator 
\Eqref{Eq:perturbation:operator} can be used not only to determine the physical 
length scales of the Ginzburg-Landau theory, but also to obtain the second 
critical field $\Hc{2}$. This is presented as a separate discussion in 
Appendix~\ref{App:Hc2}.

Finally, the length scales are given by finding the eigenstates of 
\Eqref{Eq:perturbation:operator}. More precisely, the eigenvalues $m_a^2$ 
of the (symmetric) mass matrix ${\cal M}^2$, whose elements are given in 
equations \Eqref{Eq:perturbation:1} and \Eqref{Eq:perturbation:2}, are the 
(squared) masses of the elementary excitations. The corresponding coherence 
lengths are the inverse (eigen)masses: $\xi_{a}=1/\sqrt{m_a^2}$ (and $a=I,II,III$).
Similarly, the London's penetration depth of the magnetic field is the 
inverse mass of the gauge field:  $\lambda=1/m_\A$. The mass of the gauge 
field can be read from the prefactor of $\A$ in Eq.~\Eqref{Eq:Currents2}. 
That is $m^2_\A=(q\varrho)^2$, which implies that London's penetration depth 
reads as $\lambda=q\varrho$.

The theory thus comprises four elementary length scales associated with 
the different elementary perturbations of the ground state. The length 
scale associated with the gauge field excitations is the penetration depth 
$\lambda$, and the three remaining quantities are the coherence lengths 
$\xi_{a}$ (with $a=I,II,III$). They describe at which distance the system 
recovers the ground state if one applies small perturbations of different 
linear combinations of the complex fields moduli and phase differences. 
If for example, one perturbs only one gap's modulus, several modes will be 
excited since it enters several linear combinations corresponding to different 
normal modes. Therefore there will, in general, be several length scales
in the recover of the gap module from the perturbation. 

\begin{figure*}[!htb]
\hbox to \linewidth{ \hss
\includegraphics[width=0.99\linewidth]{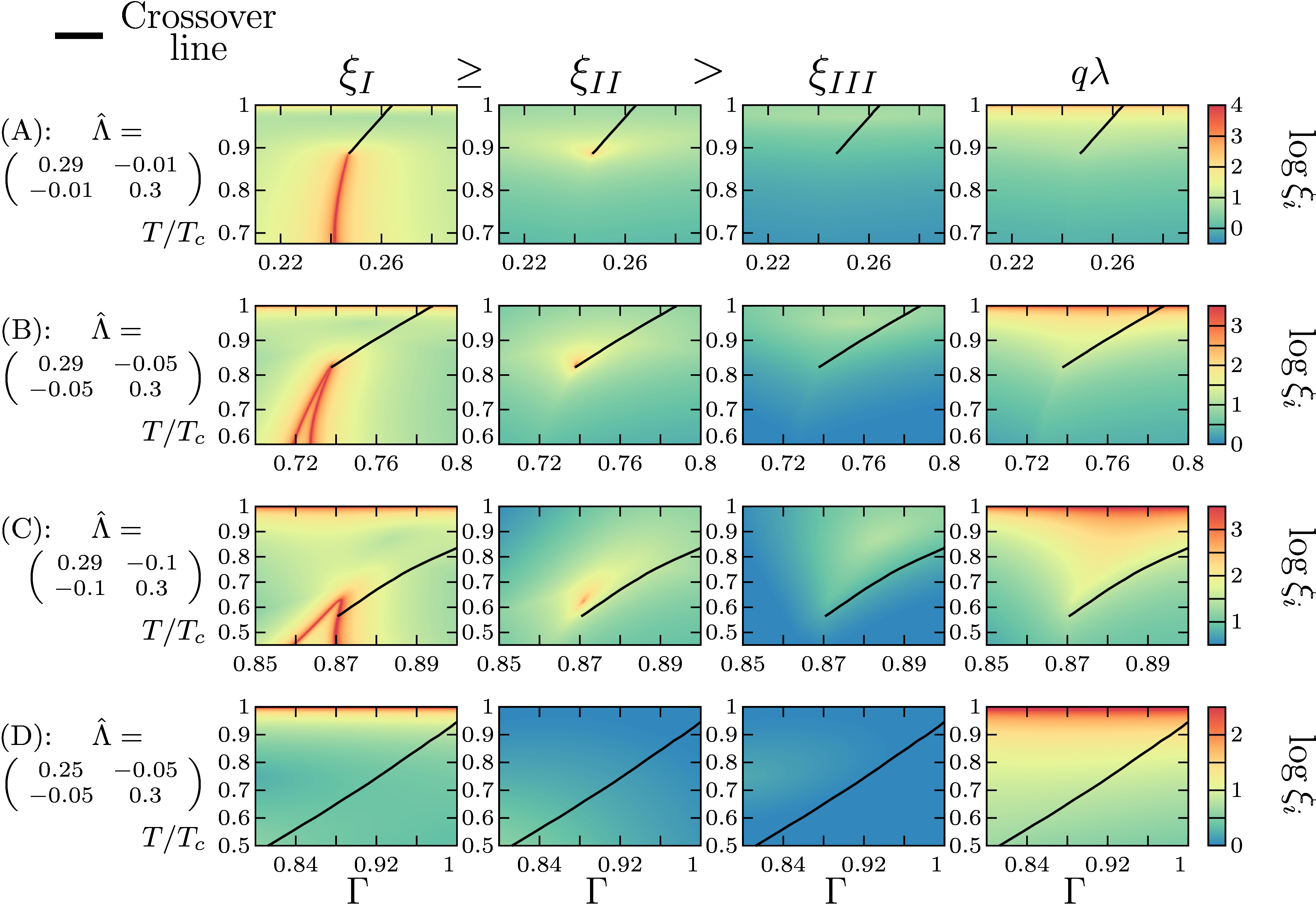}
\hss}
\caption{
%
Physical length scales corresponding to various phase diagrams of the 
Ginzburg-Landau free energy \Eqref{Eq:FreeEnergy}. These are calculated 
from the eigenvalue problem \Eqref{Eq:perturbation:operator}.
From left to right, the panels in the different columns show the three coherence
lengths   $\xi_i$ and the penetration depth $\lambda$, as functions 
of the temperature and interband scattering $\Gamma$. 
The lines correspond to different values of the coupling matrix 
$\hat{\lambda}$ that are displayed in \Figref{Fig:Diagram}. Namely lines 
(A), (B), and (C) correspond to nearly degenerate bands with weak, intermediate 
and strong repulsive interband pairing interaction respectively. Panel (D) 
describes the case of intermediate band disparity with intermediate repulsive 
interband pairing interaction. The ratio of diffusivities here is set to 
$r_d=1$ and the solid black line shows the crossover between $s_\pm$ 
and $s_{++}$ states. 
The largest coherence length  ($\xi_{I}$) diverges both at $T_c$ and and the 
transition lines from $s_{++}/s_\pm$ to the time-reversal symmetry breaking 
$s+is$ state, indicating a second order phase transition. Interestingly, 
the second largest length scale $\xi_{II}$ also diverges in a single point 
of the phase diagram corresponding to the summit of the $s+is$ dome.
Note the absence of any strong features of coherence lengths
at the crossover line between $s_\pm$ 
and $s_{++}$ states.
Note also that since the minority component vanishes at the crossover line, 
this illustrates that coherence length estimates $\xi\propto 1/\Delta$ cannot 
be used in multiband systems.
}
\label{Fig:Length-scales}
\end{figure*}

Figure \ref{Fig:Length-scales} shows such length scales in the case of 
ratio of diffusivities $r_d=1$, as functions of the temperature and 
interband scattering $\Gamma$. 
First of all, as can be seen in the first and last column of 
\Figref{Fig:Length-scales}, both the largest coherence length ($\xi_{I}$) 
and the penetration depth $\lambda$ naturally diverge at $T_c$, thus signaling 
the restoration of the $\groupU{1}$ symmetry via a second order phase transition. 
The model features additional phase transition associated with the time-reversal 
symmetry breaking: from $s_{++}/s_\pm$ (that breaks $\groupU{1}$) to the 
$s+is$ state (that breaks $\groupU{1}\times\groupZ{2}$). 
If this phase transition is second order then the largest coherence length 
($\xi_{I}$) should be divergent at that line as well. Figure~\ref{Fig:Length-scales} 
shows that this is indeed the case. Similar conclusion on the order of the 
phase transitions was reached in Ref.~\onlinecite{Silaev.Garaud.ea:17} through
analysis of the effective potential of the model. Note, however, that from 
the quantities reported in Ref.~\cite{Silaev.Garaud.ea:17}, one cannot deduce 
the coherence lengths because they depend on the gradient terms.

Interestingly, the second largest coherence length $\xi_{II}$ is always finite 
except at a single point of the phase diagram that corresponds to the summit 
of the $s+is$ dome, where $\xi_{II}$ also diverges. The shortest length scale 
($\xi_{III}$) is always finite. As can be seen from the various panels in 
\Figref{Fig:Length-scales}, all length scales are finite at the crossover 
lines (denoted by the solid black line), where one of the gap vanishes.

Physical interpretations of the different coherence lengths can be deduced 
from the analysis of eigenvectors that correspond to the normal modes. First 
of all, one should emphasize that the eigenvectors of 
\Eqref{Eq:perturbation:operator} are expressed in the \emph{rotated} basis, 
and thus do not have a direct physical interpretation in terms of the original 
pairing gaps fields. Thus the eigenvectors of perturbation operator 
\Eqref{Eq:perturbation:operator} should be expressed in the original basis. 
In analogy with the perturbative expansion \Eqref{Eq:expansion} in the 
\emph{rotated} basis, the fields in the original basis are expanded in small 
perturbations around the ground state, as 
\Equation{Eq:expansion:original}{
|\Delta_i|=U_i+\epsilon\delta|\Delta_i|\,,~~
\theta_{12}=\btheta+\epsilon\delta\theta_{12}\,.
}
There $U_i$ and $\btheta$ denote the ground state while $\delta|\Delta_i|$ 
and $\delta\theta_{12}$, stand for the perturbations in the original basis, 
and $\epsilon$ is the  small parameter of the series expansion. 
The detailed expressions of the perturbations in the original basis can be 
found in the Appendix~\ref{App:Eigenvectors}.
It is also convenient to introduce the perturbations associated to the total 
($\delta|\Delta_+|$) and relative ($\delta|\Delta_-|$) density variations, 
defined as  $\delta|\Delta_\pm|=\delta|\Delta_1|\pm\delta|\Delta_2|$.

\begin{figure}[!tb]
\hbox to \linewidth{ \hss
\includegraphics[width=0.99\linewidth]{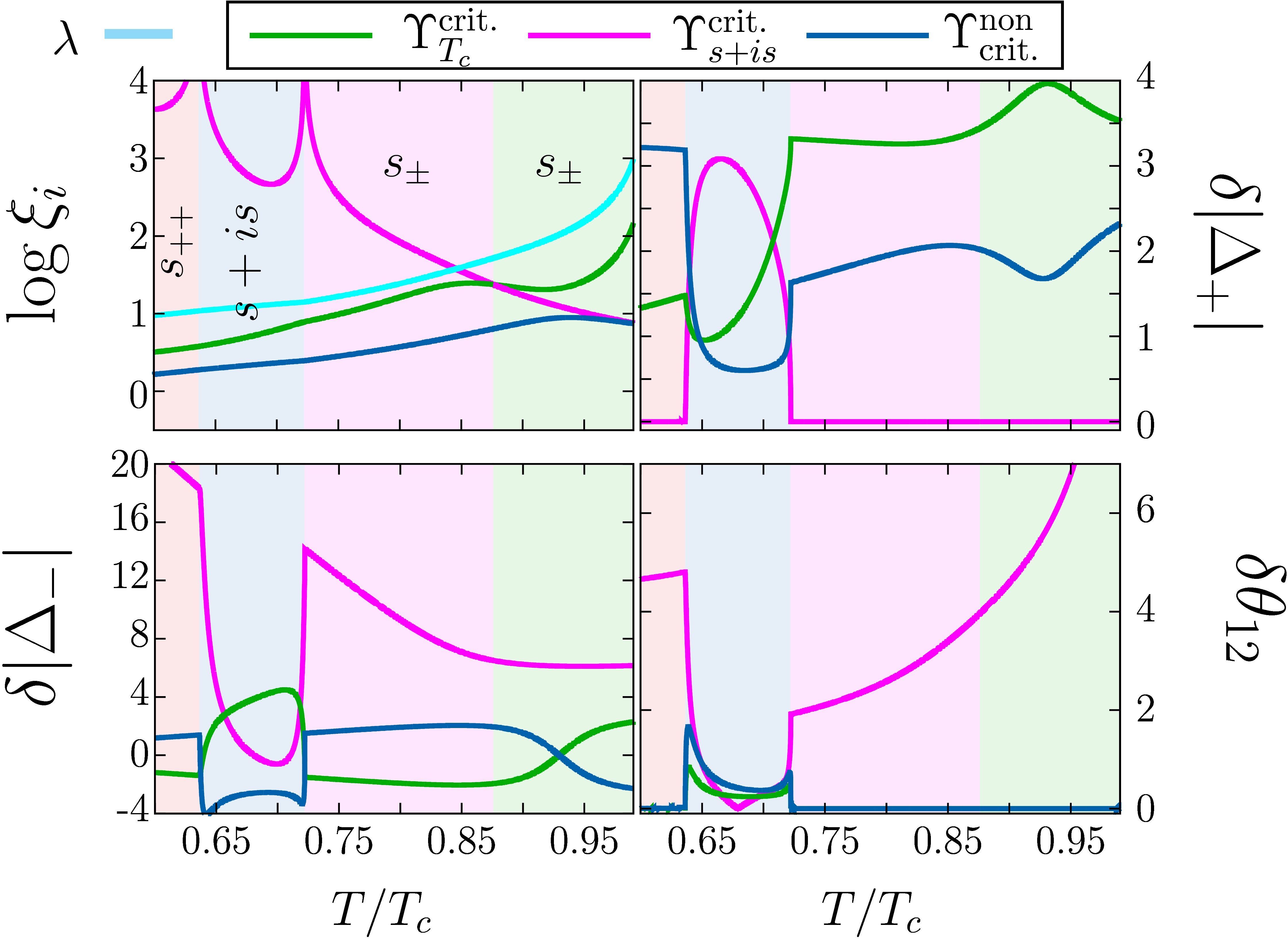}
\hss}
\caption{
Behavior of the magnetic field penetration length, the coherence lengths and 
their associated normal modes, in a dirty two-band superconductor with nearly 
degenerate bands and intermediate repulsive interband pairing interaction 
(corresponding to panel (B) of diagram \Figref{Fig:Diagram}). The first panel 
shows the penetration depth and the three coherence lengths, as functions of 
the temperature, and for a fixed interband impurity scattering $\Gamma=0.7275$. 
It thus corresponds to a vertical scan in \Figref{Fig:Diagram}(B). 
There are different normal modes that are associated with the different 
coherence lengths. The green curve refers to the mode $\Upsilon_{T_c}^\text{crit.}$, 
whose coherence length diverges at $T_c$, that is critical at the 
superconducting phase transition. The pink curve is associated with 
$\Upsilon_{s+is}^\text{crit.}$, the mode whose coherence length diverges at 
the breakdown of time-reversal symmetry. The blue curve correspond to the mode 
that is not critical.
The last three panels thus display the contributions of the different normal 
modes $\Upsilon$ to the infinitesimal perturbations $\delta|\Delta_+|$, 
$\delta|\Delta_-|$ and $\delta\theta_{12}$. Note that as $\Upsilon$ are eigenvectors 
or a linear operator, they are defined up to a normalization factor and 
only their relative contributions are meaningful.
The different background colors denote different physical regimes. The $s_{++}$ 
state is realized at lowest temperatures, then the time-reversal symmetry breaking 
$s+is$ state occurs for an intermediate temperature range and is delimited by 
two second order phase transitions with a diverging coherence length.
Finally, the $s_\pm$ is realized until $T_c$ where another second order phase 
transition occurs. Note that within the $s\pm$ phase, there are two different 
background colors. This stresses that the mode which is critical at $T_c$, 
$\Upsilon_{T_c}^\text{crit.}$, is essentially different from 
$\Upsilon_{s+is}^\text{crit.}$, the critical mode at the $s+is$ transition.
Thus the green (resp. pink) background denotes the regions where 
$\Upsilon_{T_c}^\text{crit.}$ (resp. $\Upsilon_{s+is}^\text{crit.}$) dominates.
}
\label{Fig:Length-scales-Eigenvectors}
\end{figure}

Now, given the infinitesimal perturbations \Eqref{Eq:expansion:original}, in 
terms of the perturbations $f_i$ and $\phi$ of the rotated basis, we can 
investigate the behavior of the length scales and their corresponding physical 
modes. \Figref{Fig:Length-scales-Eigenvectors} shows the length scales and the 
corresponding modes as functions of the temperature for a given interband scattering 
$\Gamma=0.7275$. This corresponds to a vertical scan in the panel (B) of diagram 
\Figref{Fig:Diagram}, going across $s_\pm$, $s+is$ and $s_{++}$ phase.
That vertical scan, covers four qualitatively different regimes. At low 
temperature, the system is in the $s_{++}$ state. The eigenmode associated with 
the largest length scale actually changes its nature during that scan. Indeed, 
the mode associated with with the divergent length scale at $T_c$ is a total 
amplitude mode, while the one that diverges at the $s+is$ transition is related 
to the relative phases. It is thus convenient to label the modes by their 
``critical" behavior. For example $\Upsilon_{T_c}^\text{crit.}$ is the mode 
that is associated with the length scale that diverges at $T_c$.
The choice, \Figref{Fig:Length-scales-Eigenvectors}, of two different background 
colors for the $s_\pm$ phase is to emphasize this fact that the mode 
$\Upsilon_{T_c}^\text{crit.}$ that dominates in the vicinity of $T_c$ has a 
completely different nature than $\Upsilon_{s+is}^\text{crit.}$ that is critical 
at the $s+is$ transition.

Interestingly, in the $s_{++}/s\pm$ phases, the mode $\Upsilon_{s+is}^\text{crit.}$ 
contributes both to relative phase and relative densities, and is decoupled from 
the total density variations. This picture produced by impurity-scattering is in 
contrast to clean two-band case \cite{leggett} where phase difference is fully 
decoupled from densities at linear level. 
Thus starting from low temperature, state in the $s_{++}$ phase, 
$\Upsilon_{s+is}^\text{crit.}$ does not contribute to the total density, but 
couples relative phase and relative density.
At a higher temperature a second order phase transition to the time-reversal 
symmetry breaking $s+is$ state occurs, signaled by the divergence of the largest 
coherence length. In the $s+is$ state, all the modes contribute to the density 
modes (total and relative) and to the relative phase excitations as well. 
Further increasing the temperature drives the system through another second order 
phase transition to the $s_\pm$ state. Importantly, when approaching $T_c$, 
the critical mode at $s+is$ transition $\Upsilon_{s+is}^\text{crit.}$ that 
was dominating becomes subdominant in favor of $\Upsilon_{T_c}^\text{crit.}$, 
a pure density (amplitude) mode that is relevant for the restoration of the 
normal state.

The results of the length scale analysis reported in Figs.~\ref{Fig:Length-scales}
and \ref{Fig:Length-scales-Eigenvectors} are performed for equal electron 
diffusivities in the different bands ($r_d=1$). Varying the relative diffusion 
constant
alter the results only quantitatively, while the overall picture described above 
remains qualitatively the same. Quantitative detail on the influence of the
relative diffusion constant
$r_d$ on the length scales and on the upper critical field are reported 
in Appendix~\ref{App:rd}.
The analysis above shows that, at the linear level, the normal modes of a dirty 
two-band superconductor always couple the density and the relative phase excitations. 
Therefore such system does not feature a phase-only Leggett's mode. This has to be 
contrasted with the case of similar but clean two-band system \cite{Silaev.Babaev:11,
Silaev.Babaev:12}, where the Leggett's modes and density modes always decouple.

Complicated variations of the coherence lengths in the dirty case, as well as the 
existence of diverging coherence lengths, are consequences of competing $s_\pm$ 
and $s_{++}$ and the $s+is$ states. They should have physical manifestations through 
the various responses that involve spatial or dynamical variations of the fields. 
Although their detailed analysis is beyond the scope of the current paper, 
we mention a few phenomena that can arise as a consequence of the rich interplay 
of the normal modes and their corresponding length scales.
The above calculations do not consider dynamics but it demonstrates the existence 
of massless and soft dynamical modes that can be directly probed in experiment 
\cite{Blumberg.Mialitsin.ea:07}.The mixed modes also dictate nontrivial 
thermoelectric properties \cite{Garaud.Silaev.ea:16,Silaev.Garaud.ea:15} 
and their softening manifests itself in anomalies of flux flow viscosity 
\cite{Silaev.Babaev:13}. Likewise by the same mechanism the mode mixing 
produces nontrivial magnetic signatures of impurities \cite{Lin.Maiti.ea:16,
Maiti.Sigrist.ea:15}, we discuss this in more detail below. 
Another interesting feature, that follows from the fact that one length scale 
diverges near the transition to the $s+is$ state, is that it can result in a 
particular length scale hierarchy where the magnetic field penetration length
becomes an intermediate length scale. In the next section we consider implications 
of such a length scale hierarchy on vortex matter, and in particular illustrate 
that some behavior that can be deduced from the length scale analysis, 
actually survive beyond the linear regime.

\section{Vortices in the vicinity of the \texorpdfstring{$s+is$}{s+is} region, 
physics beyond the linear regime}
\label{Sec:Vortices}

Here we discuss the physical properties associated with the topological 
excitations of dirty two-band superconductors, especially focusing on the 
possible consequences of the presence of the $s+is$ critical line on the phase 
diagram. 
We thus construct vortex solution by numerically minimizing the free energy 
\Eqref{Eq:FreeEnergy}. The physical degrees of freedom $\Delta_1$, $\Delta_2$ 
and $\A$ are discretized using finite-element formulation \cite{Hecht:12}, and 
the free energy is minimized using a non-linear conjugate gradient algorithm.
To construct vortex solutions the minimization procedure is started with an 
initial configuration, in which both components $\Delta_1$ and $\Delta_2 $ have 
the same vorticity. This initial vorticity specify the number of vortices that  
originally seeded in the numerical grid. The minimization procedure leads, 
after convergence of the algorithm, to a vortex configuration that carries the 
number of flux quanta that was specified by the initial phase winding. Note that 
the numerical grid has to be chosen much larger than the vortex configurations 
that are constructed. This is important to ensure that vortex matter do not 
interact with the domain boundaries and thus that the obtained configurations 
are not artifacts of boundary interactions. In particular, in the results that 
are displayed below, the numerical grid is larger than the displayed region 
which are close up views of the regions carrying vortices 
\cite{
[{For further details on the numerical methods employed here see, for example, 
related discussion in: }]
[{}] Garaud.Babaev.ea:16}. 
Note also that here we are interested in the physical properties of the vortex 
matter, such as intervortex forces, rather than magnetization process. This is 
why vortices are constructed here in zero external field, and seeded by the 
initial guess. By contrast in an external field, the vortex matter is not only 
subjected to its own intervortex interactions, but also to the interaction with 
Meissner currents, surface  barriers, etc.

The determination of the length scales of the theory described in the previous
section, relies on the linearization of the theory around the ground state. 
This analysis is thus relevant in the asymptotic regions that are far away from 
the vortex cores. As a result, the long-range intervortex interactions (that is, 
the interactions in the asymptotic region where the linear theory holds), are 
described in terms of the coherence lengths associated with the normal modes 
of the system [the solutions of the eigenproblem \Eqref{Eq:perturbation:operator}]. 
Their nontrivial evolution and mixing across the phase diagram indicates the 
possible realization of nontrivial intervortex physics beyond the linear regime 
and thus a likely unusual magnetic magnetic response of the system.
Because the dirty two-band superconductors described here feature a critical 
line that segregates the $s+is$ state from the other $s$-wave states, one 
coherence should diverge in the vicinity of that transition. Thus varying the
temperature can drive the system from an $s$-wave state through the second 
order phase transition to the time-reversal symmetry-breaking $s+is$ state.
Interestingly, as stressed in \Figref{Fig:Length-scales-Eigenvectors}, for some 
fixed values of the impurity scattering rates, there can be two successive second 
order phase transitions: one from the $s_\pm\to s+is$ followed by a second 
$s+is\to s_{++}$ transition at lower temperature.
As illustrated in \Figref{Fig:Length-scales-Eigenvectors}, it immediately follows 
that a temperature-driven phase transition to the $s+is$ state goes along with 
the divergence of the largest coherence length at the transition line, while all 
other length scales, including the magnetic field's penetration depth, remain finite.

Since the other length scales, including the magnetic field's penetration length 
are finite at this transition, there are only two possible hierarchies of the 
length scales near that transition: (i) all coherence lengths are larger than 
$\lambda$ (which is a type-1 behavior), (ii) $\xi_I>\lambda$ but $\lambda$ is 
larger than some of the other coherence lengths. Since intervortex interactions 
are related to the long-range asymptotics, such a hierarchy of the length scales 
suggests long-range attractive, short-range repulsive intervortex forces. This 
regime was earlier termed ``type-1.5" \cite{Moshchalkov.Menghini.ea:09} while 
associated phase separation was termed ``semi-Meissner" state \cite{Babaev.Speight:05}). 
As emphasized above, in dirty two-band superconductor, 
the normal mode with the largest coherence length typically mixes density modes 
and phase-difference mode. This implies that, in the vicinity of the $s+is$ 
transition, vortices feature a long-range tail of density suppression. This 
results in long-range attractive intervortex forces (dominated by the core-core 
interactions). On the other hand, at intermediate scales specified by the magnetic 
field's penetration depth, the interactions are dominated by current-current 
interactions which are repulsive. The long-range intervortex interacting 
potential predicted by the linear theory can be expressed as a combination
of modified Bessel functions of the second kind $K_0$, as:
\begin{equation}
 U(r)=-C_{\lambda}^2 K_0\left(\frac{r}{\lambda}\right)+\sum_{i=I,II,III} \left[
     C_i^2  K_0\left(\frac{r}{\xi_i}\right)\right] \,.
\label{Eq:Asymptotics}
\end{equation}  
The coefficients $C_\lambda$ and $C_i$ depend on the eigenstates of the perturbation 
operator (the normal modes) and on nonlinearities.
%
%
\begin{figure*}[!htb]
\hbox to \linewidth{ \hss
\includegraphics[width=0.575\linewidth]{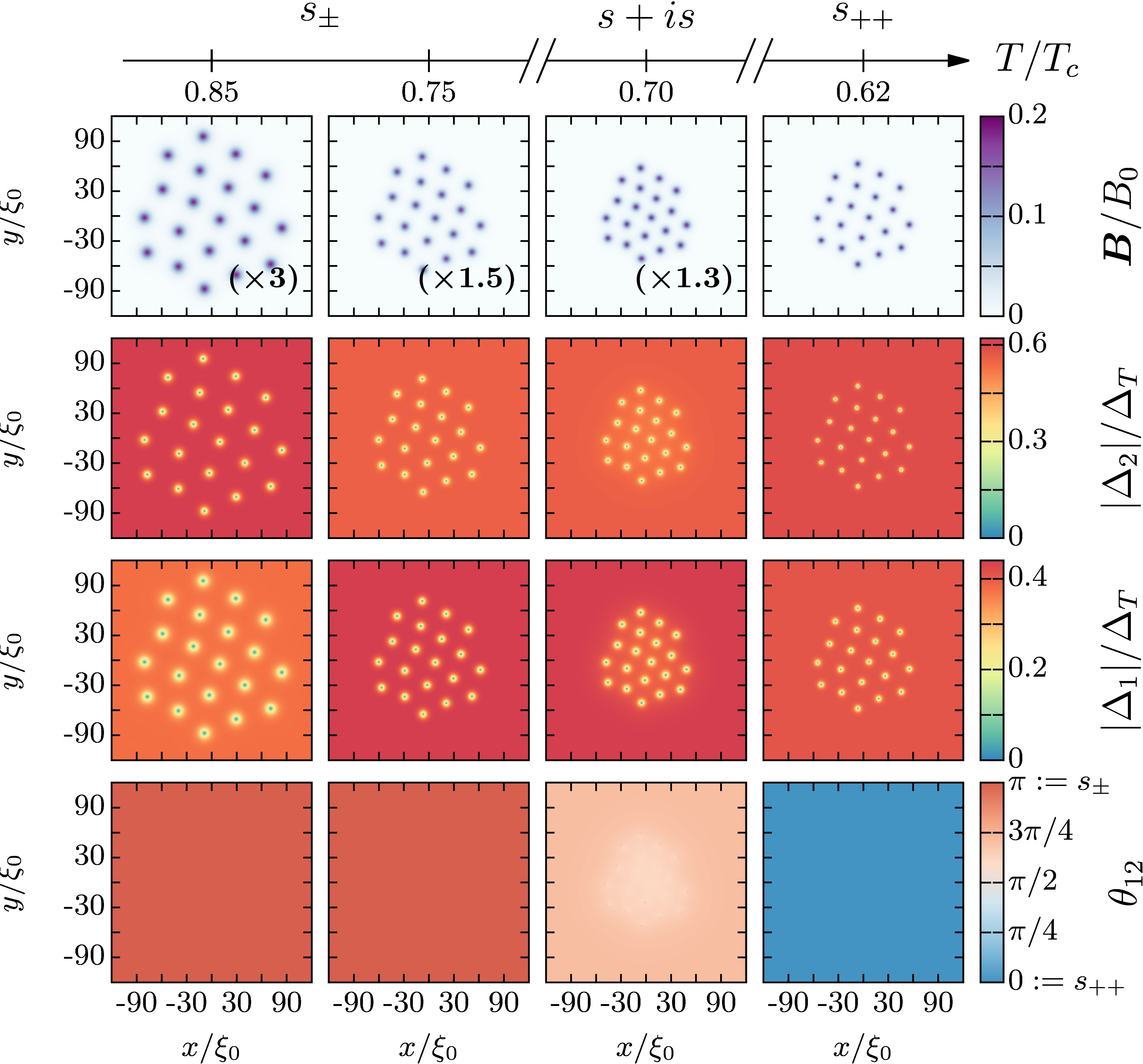}
\hspace{0.035\linewidth}
\includegraphics[width=0.285\linewidth]{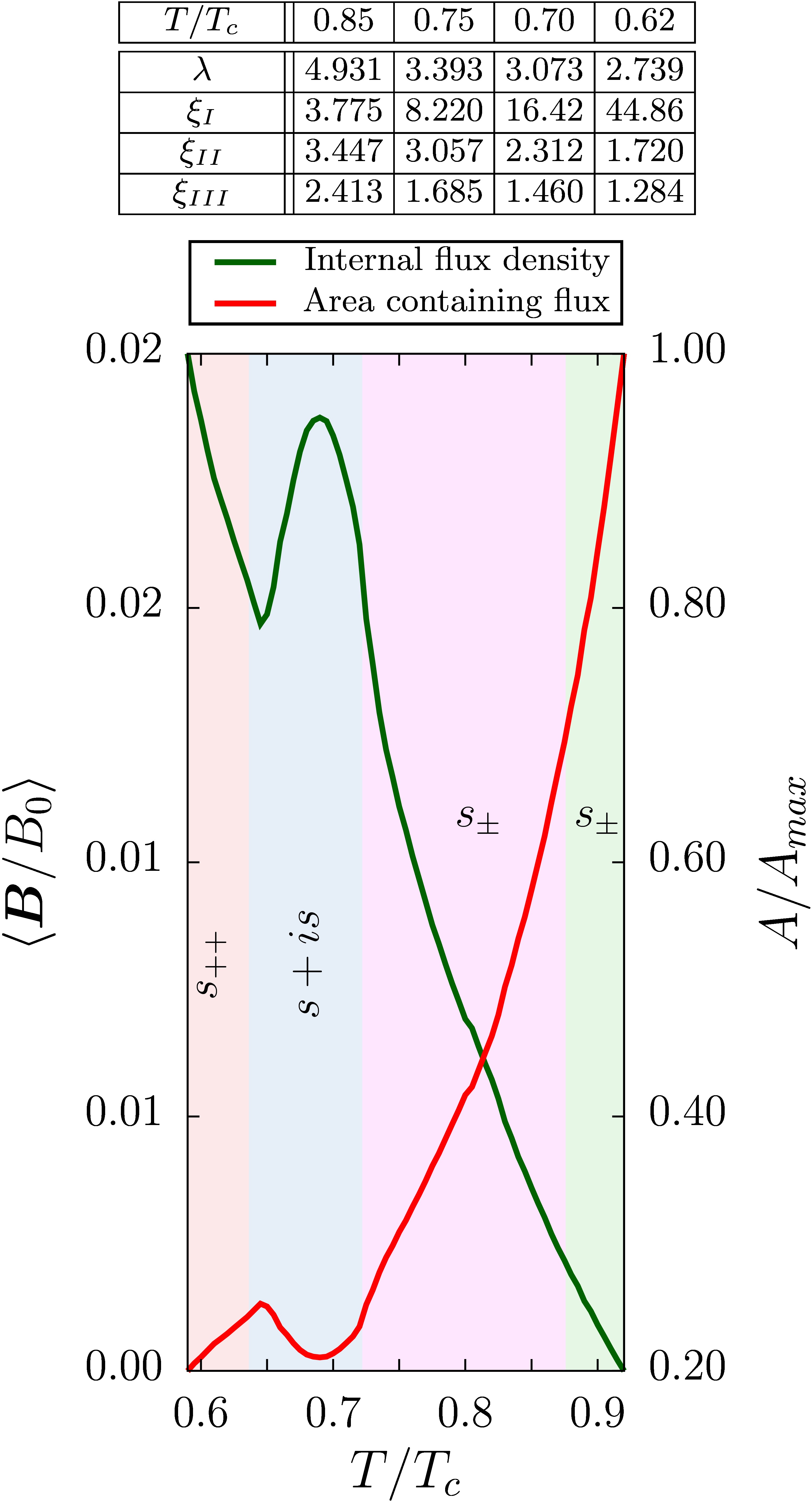}
\hss}
\caption{
Evolution of bound state of 20 vortices during a cooling procedure like that 
displayed in \Figref{Fig:Length-scales-Eigenvectors}. Apart from the temperature, 
all other paramaters are fixed: the interband scattering rate is $\Gamma=0.7275$, 
the gauge coupling constant is $q=0.5$, and electronic diffusivities are equal 
($r_d=1$).
On the left panel, the displayed quantities on the different lines are respectively 
the magnetic field $\B$, the amplitudes of the gap in majority ($|\Delta_2|$) 
and minority ($|\Delta_1|$) components. The last line shows the relative phase 
$\theta_{12}$ from which the ground state phase can be directly read.
Different vortex configurations, at different temperatures, can be read from the 
various columns displayed on the left panel. The first column shows a type-2 
regime where the largest length scale is the penetration depth and thus the 
repulsion forces dominate. The other columns, on the other hand, show the typical 
realization of a type-1.5 regime. There, due to the proximity with the 
$s_{++}/s_\pm\to s+is$ transition, the largest coherence length increases and 
this triggers the long-range attractive forces resulting in the formation of a 
compact cluster of vortices. 
The right panel shows the flux-carrying area (defined as the area of the region 
where the magnetic fields is above some threshold $\delta = 0.005 B_{max}$), 
and the internal mean magnetic flux density in the flux-carrying region. 
The internal flux density shows a strong peak 
where the attractive intervortex forces are strongest and thus the
clusters are the most compact. Here this peak is near the $s+is$ transition.
The numerical values of the various length scales corresponding to the different 
regimes displayed on the left panel are shown in the top right table.
}
\label{Fig:Cluster20}
\end{figure*}
%
Thus, in the vicinity of the second order phase transition to the time-reversal 
symmetry breaking $s+is$ state, the interplay between the long-range attraction 
driven by the core-core interactions, and the short-range repulsion due to the 
current-current repulsion, yields non-monotonic intervortex forces (cf. with 
calculations in different two-band models \cite{Babaev.Speight:05,
Babaev.Carlstrom.ea:10,Carlstrom.Babaev.ea:11,Silaev.Babaev:11}.)

Such forces can promote the formation of a bound-state of vortices. In such 
a bound-state, the distance separating the vortices does not directly follow 
from linearized theory, but is determined by full nonlinear theory. As a result, 
since all the parameters of the Ginzburg-Landau model \Eqref{Eq:FreeEnergy} are 
temperature dependent (see the exact formulas of the coefficients in Appendix 
\ref{App:Coefficients}), it is quite expectable that if vortex bound-state are 
formed in the full nonlinear model, their typical size should also be temperature 
dependent.

Below, we present the results of such an analysis of the full nonlinear response 
in the Ginzburg-Landau model. Using the numerical procedure described earlier in 
this section, we systematically construct sets of several vortices for fixed 
impurity scattering rates and decreasing temperatures similar to the parameter 
set investigated in \Figref{Fig:Length-scales-Eigenvectors} (thus corresponding 
to a vertical scan in the phase diagrams \Figref{Fig:Diagram}). 
The situation that is considered now mimics a dilute  groups of vortices that 
form in a field-cooled sample in fields far from upper critical magnetic field. 
Figure~\ref{Fig:Cluster20} displays the behavior of a set of $20$ vortices at 
different temperatures. The selected temperatures are representatives of the 
various phases shown in \Figref{Fig:Length-scales-Eigenvectors}.

Depending on the regime, when starting from an initial set of $20$ vortices, 
the numerical procedure leads after convergence to a characteristic picture 
corresponding to either a type-2 regime or to vortex clusters that are typically 
realized in the type-1.5 regime. In the vicinity of the superconducting transition, 
as illustrated by the configuration in the first column of \Figref{Fig:Cluster20}, 
which is close to $T_c$ and deep in the $s_{\pm}$ region, the vortex configuration 
is typical of a type-2 regime. Note that type-2 regime theoretically implies an 
infinite vortex separation, but the strength of the repulsion decays exponentially 
with the separation. So for all practical purposes, the repulsion between vortices 
ends when the strength becomes smaller than the numerical accuracy: that is similar 
to experimental situation of remnant vorticity where intervortex repulsion or 
vortex-boundary interaction is too small to reach truly lowest energy state. 

Upon decreasing the temperature, the largest coherence length $\xi_I$ increases 
rapidly, as the system gets closer to the transition to the $s+is$ state. This 
triggers the expected long-range attractive mode which leads to the formation 
of a vortex cluster. 
The last three columns on the left panel of \Figref{Fig:Cluster20} correspond, at 
the linear level to type-1.5 regimes. In other words, as can be read from the values 
of the length scales on \Figref{Fig:Length-scales}, the penetration depth there, 
is and intermediate length scale. The corresponding regimes in the last three 
columns of \Figref{Fig:Cluster20} show that, in the nonlinear regime, vortices 
aggregate in a cluster. 
As can be seen from the two central columns of \Figref{Fig:Cluster20}, the vortex 
coalescence occurs near the phase transition to the $s+is$ state, and the most 
compact cluster forms in the $s+is$ phase (this can seen from the third column). 
Further decreasing of the temperature drives the system through another 
second-order phase transition to the $s_{++}$ state. While moving away from 
criticality, the largest coherence length shortens. Correspondingly, the range of 
attractive interaction also shortens, and the attractive forces weaken. Eventually, 
repulsive forces will become dominant again, and the set of vortices will fall 
back in a type-2 regime. Observe that \Figref{Fig:Cluster20} clearly shows that 
the scale of strong suppression of gaps in the vortex core is not directly related 
to coherence lengths.

For a system with non-monotonic interactions, standard kinetic mechanisms (see 
for example \cite{Sethna}) leads to different patterns of phase coexistence.
In the case studied here the vortex clusters coexist with domains of Meissner state. 
The temperature dependence of intervortex forces opens up a possibility to discriminate 
the effect described here from a phase separation originating in vortex pinning. 
The formation of vortex cluster can be probed by the direct vortex visualization 
techniques such as magnetooptics, scanning Hall, and scanning SQUID probes. 
This could also be experimentally probed for example in muon-spin rotation 
measurements ($\mu SR$) like the ones conducted in Ref.~\cite{Ray.Gibbs.ea:14}. 
That is, when 
$\mu SR$ detects a phase separation into vortex clusters and Meissner domains, 
the above considered contraction of vortex cluster when the temperature is lowered, 
should result in a local increase of magnetic field: quantity that, again, can 
be extracted from $\mu SR$ data \cite{Ray.Gibbs.ea:14}. In order to connect the 
effect of vortex clusterization with this experimentally measurable quantity, 
we calculate the local mean magnetic flux density for vortex cluster as 
the system is cooled down. The evolution of the mean magnetic flux density of a 
cluster during the cool-down process is displayed on the right panel of 
\Figref{Fig:Cluster20}.
There is a strong peak in the magnetic flux density in correspondence to an 
increase of the vortex binding forces near the $s+is$ transition. 

\begin{figure}[!htb]
\hbox to \linewidth{ \hss
\includegraphics[width=0.99\linewidth]{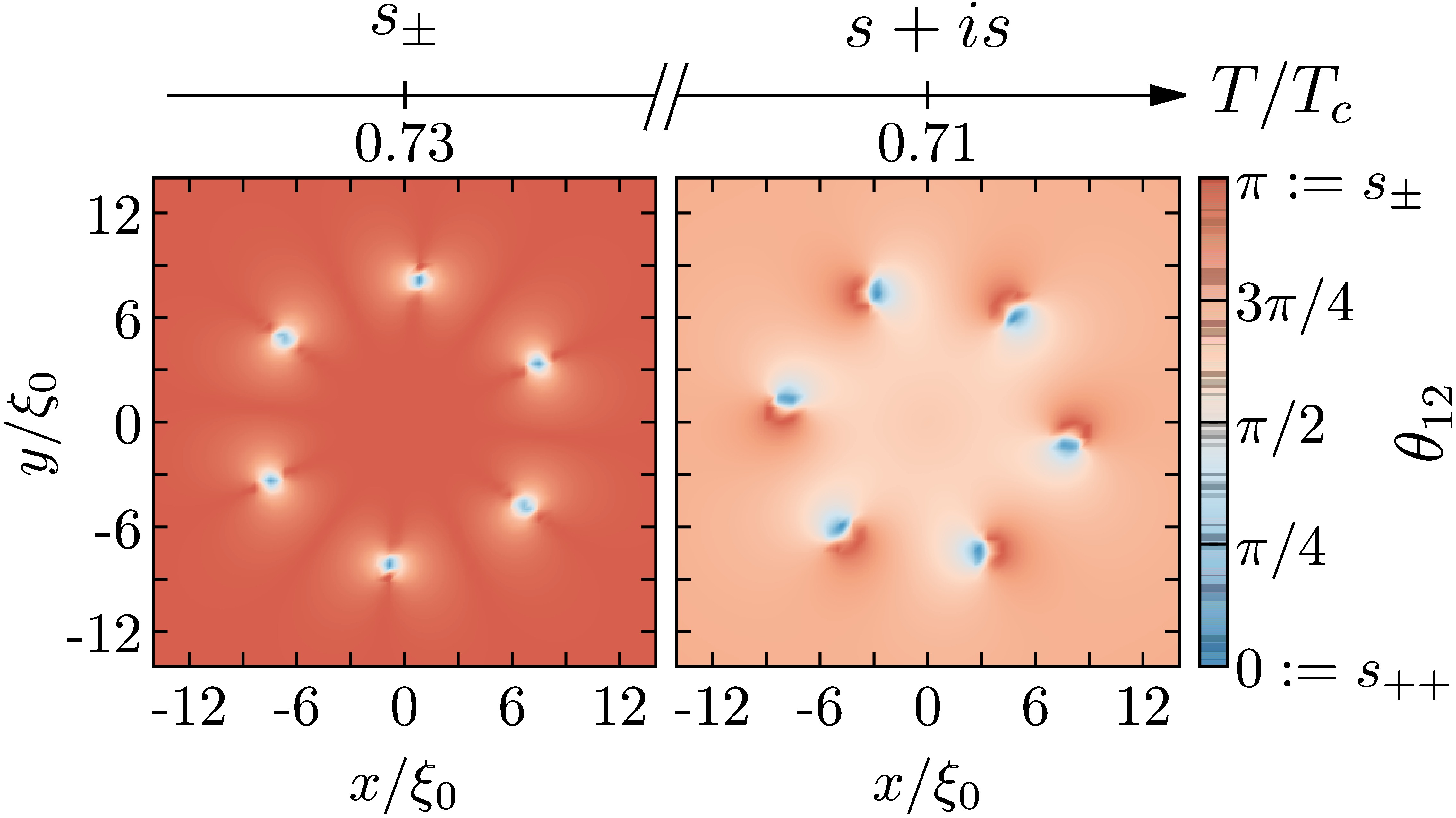}
 \hss}
\caption{
Splitting of composite integer flux quanta vortices into fractional vortices,  
at the boundary of clusters formed due to the long range attractive forces near 
the $s+is$ transition. Here a cluster of seven vortices for the same parameters 
as in \Figref{Fig:Cluster20} besides $q=0.8$, both in the $s_{\pm}$ state 
(first column) and $s+is$ (second column). 
The relative phase shows the splitting of vortices at the boundary of 
the cluster.
}
\label{Fig:Skyrmions}
\end{figure}

The appearance of this kind of signal assumes phase separation due to kinetic 
reasons. However similar signal should also be expected for dilute vortex lattices 
that can contract due to emerging attractive forces as well. The strength of the 
effect will also depend on the magnetic penetration lengths: the longer is $\lambda$, 
the weaker are the intervortex attractive forces. On the quantitative side: an 
interesting feature is the the vortex clusterization can start very far away from 
the $s+is$ phase transition. This is fully consistent with linear analysis where 
we find, in \Figref{Fig:Length-scales-Eigenvectors}, a broad region of increased 
largest coherence length associated with the critical mode. Therefore even if the 
$s+is$ phase occupies an unobservably small domain on the phase diagram the soft 
modes implied by that criticality exist and modify magnetic response in a wide 
range of parameters.

Having established vortex clustering due to existence of a critical mode, we 
briefly discuss a few of structural features of vortex clusters. Detailed study 
of the vortex cluster structure is beyond the scope of this paper and is perhaps 
a fruitful direction of application of methods developed in research on filament 
bundles. Yet, it should be emphasized that indeed the vortex clusters are not a 
simple superposition of single vortex solutions. Correspondingly the equation 
\Eqref{Eq:Asymptotics}, that is based on the linear theory with the assumption 
of axially symmetric composite vortices, can received nonlinear corrections. 
One of the possible nonlinear effect is that clusters can exhibit disintegration 
of the composite character of vortices: namely a small splitting of the vortex 
cores in the different components near clusters boundary. This behavior can clearly 
be seen in the numerical solutions shown in \Figref{Fig:Skyrmions} for a cluster 
of seven vortices. 
As can be seen in \Figref{Fig:Skyrmions}, clearly the phenomena of vortex 
splitting occurs at the boundary, while the inner vortex sitting at the center of 
the cluster shows no splitting. Such a splitting of vortex cores at the boundary 
of clusters excites a mode that is not present in the interaction between single 
vortices, but that should contribute to the inter-cluster interactions. 

It should finally be stressed the coherence lengths cannot be related in a simple 
way to overall vortex core sizes, because of the nonlinear nature of the coupled 
system. It can clearly be seen in \Figref{Fig:Cluster20} that although one of the 
length scales diverges in the vicinity of the $s+is$ transition, this has a 
relatively little influence on the size of substantial density suppression in vortex 
cores (but dramatically affects the long-range weak density suppression and thus
the intervortex interactions). 

\section{Effects of spatial gradients in impurity density}

The previous sections describe the effect of impurities on the phase diagram of 
dirty two-band superconductors and, in particular, how it affects the different 
length scales. We further demonstrated that, in the vicinity of the impurity 
induced second order phase transition associated with the spontaneous breakdown 
of the time-reversal symmetry, it results in non-monotonic intervortex forces 
that can lead to the formation of vortex clusters whose typical signature 
can in principle be probed in $\mu$SR measurements. 
%
%
\begin{figure}[!htb]
\hbox to \linewidth{ \hss
\includegraphics[width=0.99\linewidth]{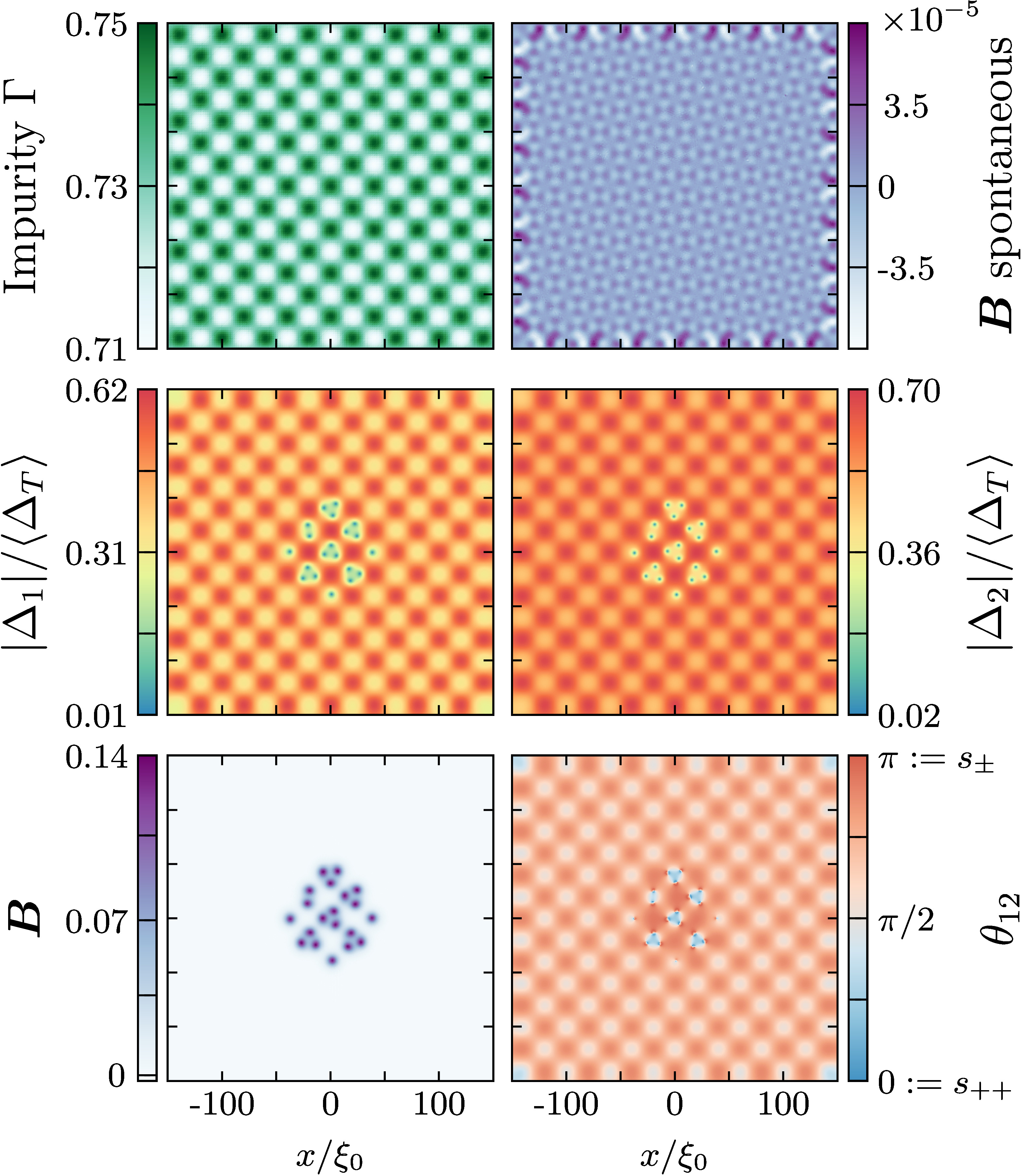}
 \hss}
\caption{
Appearance of spontaneous magnetic fields and fragmentation of vortex cluster 
due to small periodic modulation of the impurity scattering rate $\Gamma$. 
All the microscopic parameters are the same as in \Figref{Fig:Cluster20}. 
The temperature is $T/T_c =0.7$, while the impurity scattering strength is 
modulated sinusoidally with amplitudes $\Gamma=0.7275 \pm 0.02$. 
The panels on the top row show the modulated impurity and the induced magnetic 
field, in the absence of vortices. The middle line displays the amplitudes of 
the gap in minority ($|\Delta_1 |$) and majority components ($|\Delta_2 |$) 
of the fragmented of vortex cluster. The bottom row shows the corresponding 
magnetic field $\B$ and the relative phase $\theta_{12}$.
Modulation of impurity provides a pinning landscape that favors placing vortices 
where the impurity is increased, and this tends to break cluster into smaller 
clusters, as compared to \Figref{Fig:Cluster20}. 
Nonetheless the clearly large areas of gap suppression indicates the presence 
of attractive intervortex forces. In addition the modulation of $\Gamma$ produces 
a small spontaneous magnetic field in the $s+is$ state as can be seen from the 
second top panel. Note however that the spontaneous field is several orders of 
magnitude smaller than that of vortices, and it is typically dominated by the 
vortex background.
}
\label{Fig:Impurity}
\end{figure}
%
%
The discussion so far focused on the case of a spatially uniform distribution 
of impurity density. It is instructive to consider one more example, where the 
impurity density is not uniformly distributed in space. Spatially varying impurity 
will result in an inhomogeneous superconducting state which will feature gradients both of densities and relative phases of the 
field components as a consequence of 
mode mixing. This in turn can produce spontaneous magnetic fields.
Spontaneous magnetic fields were indeed shown to occur in various models of 
$s+is$ states for different kind of inhomogeneities such as impurities and 
domain walls and their combinations \cite{Garaud.Babaev:14,
Maiti.Sigrist.ea:15,Lin.Maiti.ea:16,Vadimov.Silaev:18}, and impurity-like inhomogeneities 
produced by the local heating of a superconductor \cite{Silaev.Garaud.ea:15,
Garaud.Silaev.ea:16}.
In this section we report that spontaneous magnetic fields arise when there are 
gradients of impurity density. 
The magnetic field discussed in the previous section will be 
superimposed with the spontaneous magnetic field. 
This can have direct implications for the vortex 
states previously considered, since both the inhomogeneities and the spontaneous 
magnetic fields induced by them can provide a pinning landscape for vortices.

To illustrate that inhomogeneities can indeed yield spontaneous magnetic fields, 
we consider an idealized situation of a sinusoidal modulation of the impurity 
scattering rates $\Gamma=0.7275 \pm 0.02$, where the period of the modulation is 
of the order of the size of a vortex.
Figure~\ref{Fig:Impurity} shows this situation, where spontaneous magnetic 
field appears due to the modulation of impurities. Moreover, it shows that it 
can also substantially affect vortex structures such as the clusters previously 
reported in \Figref{Fig:Cluster20}. Indeed, in this example, inhomogeneities of 
the impurity scattering rates clearly result in the fragmentation of the vortex 
cluster into smaller clusters. In other words, inhomogeneities induce a pinning 
landscape due to gap modulations and due to appearance of spontaneous multipolar 
magnetic field, that affects the structure of vortex clusters.

The spontaneous magnetic field that arises exclusively due to inhomogeneities 
(i.e. without vortices) is displayed in the second panel of the first row of 
\Figref{Fig:Impurity}, and it is maximal where the modulation of $\Gamma$ has 
its larger gradients. The spontaneous magnetic field is spatially alternating 
and its total flux across  the sample is zero in the absence of vortices. 
Note that there is an enhanced spontaneous field near the boundaries that is 
dipole-like, rather than quadrupole-like in the bulk. This is not a generic 
property, but rather a consequence of the particular choice of the modulation 
here that has stronger gradients whose contribution is less compensated at the 
boundary. 

The presence of spatially inhomogeneous distribution of impurities can thus have 
experimental manifestations that can, for example, be detected in a zero field 
by $\mu SR$ and scanning SQUID experiments. Moreover, as can be seen from the 
other panels in \Figref{Fig:Impurity}, besides the spontaneous magnetic fields, 
the inhomogeneities can also act as a pinning landscape for vortices that will 
alter the structure of the clusters discussed in the previous section. Such a 
cluster fragmentation due to pinning should result in a reduction of the $\mu$SR 
signatures discussed in the previous section, in the case of homogeneous systems.

\section{Conclusion}
\label{Sec:Conclusion}

In conclusion, in this work we studied the properties of dirty two-band 
superconductors with repulsive interband interaction. We used the microscopically 
derived Ginzburg-Landau theory, to give qualitatively consistent solutions with 
Usadel model, not too far from superconducting $T_c$.
We investigated the normal modes, and their corresponding coherence lengths. 
The normal modes of dirty systems are much more complex than those in clean 
two-band cases due to the frustration between various interband interaction terms.
One of the new features is the mixing of the Leggett mode with the density modes 
that occurs even without time-reversal symmetry breaking. 
An important property of the dirty two-band superconductors is the presence of a 
region of $s+is$ state on the phase diagram. This $s+is$ domain is surrounded by 
a line of second order phase transition that dictates the existence of a soft mode 
and an infinite disparity of coherence lengths. 
A striking feature is that the disparity of the coherence lengths and relatively 
soft mode persists for a wide range of coupling constants and temperatures, even 
if the domain of the $s+is$ phase is too small to be directly resolvable in 
experiment. This, makes such systems very different from the clean two-band 
$s$-wave case \cite{Silaev.Babaev:11}. 
This should also have consequences for various properties associated with the 
static and dynamic fluctuations. We focused here on consequences to the vortex 
physics and demonstrated the existence of a region where the hierarchy of the 
length scales is such that the penetration depth becomes an intermediate length 
scale (the so-called type-1.5 regime). This leads to long-range attractive and 
short-range repulsive intervortex forces, leading to formation of vortex bound 
states or clusters. These clusters, due to the temperature dependence of the 
mean magnetic field's density, have specific signatures that can be discriminated 
from other mechanisms also responsible of cluster formation. This should be 
experimentally measureable in muon-spin-rotation experiments. 
Note finally that qualitatively similar features should also be expected in 
other realizations of $s+is$ states \cite{Ng.Nagaosa:09,Stanev.Tesanovic:10,
Carlstrom.Garaud.ea:11a,Maiti.Chubukov:13,Boeker.Volkov.ea:17,Grinenko.Materne.ea:17}.

\begin{acknowledgments}
The work was supported by the Swedish Research Council Grants
No. 642-2013-7837, VR2016-06122 and Goran Gustafsson Foundation 
for Research in Natural Sciences and Medicine. The computations 
were performed on resources provided by the Swedish National 
Infrastructure for Computing (SNIC) at National Supercomputer 
Center at Link\"oping, Sweden.
\end{acknowledgments}


\clearpage
\appendix
\renewcommand{\theequation}{\Alph{section}\arabic{equation}}

\section{Ginzburg-Landau coefficients}
\label{App:Coefficients}

There, the coefficients of the Ginzburg-Landau functional $a_{ij}$, $b_{ij}$, 
$c_{ij}$ and $k_{ij}$ can be calculated from the inputs $\lambda_{ij}$, $T$ 
and $\Gamma$ of the microscopic self-consistency equation. $N_i$ are the 
densities of states and $D_i$ the electron diffusivities. First, the 
coefficients of gradient terms are given by \cite{Garaud.Silaev.ea:17a}
\SubAlign{Eq:GLparameters:K}{
 k_{ii} &= 2\pi T N_i \sum_{n=0}^{N_d} 
 \frac{ D_i (\omega_n + \gamma_{ji})^2 + \gamma_{ij}\gamma_{ji} D_j }
 {\omega^2_n(\omega_n+\gamma_{ij}+\gamma_{ji})^2} \\
 k_{ij} &= 2\pi T N_i\gamma_{ij} \sum_{n=0}^{N_d} 
 \frac{ D_i (\omega_n + \gamma_{ji}) + D_j (\omega_n + \gamma_{ij}) }
 {\omega^2_n(\omega_n+\gamma_{ij}+\gamma_{ji})^2}	\,,
}
with $j\neq i$. The coefficients of the potential terms can be found 
for example from Ref.~\onlinecite{Stanev.Koshelev:14} and they read as 
\SubAlign{Eq:GLparameters:A}{
a_{ii}&= \frac{N_i\lambda_{jj}}{\mathrm{det}(\hat{\lambda})}
-2\pi T \sum_{n=0}^{N_d}
\frac{(\omega_n+\gamma_{ji}) N_i}{\omega_n(\omega_n+\gamma_{ij}+\gamma_{ji})}
\,, \\
a_{ij}&=-\frac{N_i\lambda_{ij}}{\mathrm{det}(\hat{\lambda})}
-2\pi T \sum_{n=0}^{N_d}
\frac{ \gamma_{ij} N_i}{\omega_n(\omega_n+\gamma_{ij}+\gamma_{ji})}
\,. 
}
The other parameters read as 
\SubAlign{Eq:GLparameters:B}{
b_{ii}&= \pi T N_i\sum_{n=0}^{N_d}
\frac{(\omega_n+\gamma_{ji})^4 }{\omega_n^3(\omega_n+\gamma_{ij}+\gamma_{ji})^4}
\\ &~+
\pi T N_i\sum_{n=0}^{N_d}
\frac{\gamma_{ij}(\omega_n+\gamma_{ji})
(\omega_n^2+3\omega_n\gamma_{ji}+\gamma_{ji}^2) }
{\omega_n^3(\omega_n+\gamma_{ij}+\gamma_{ji})^4}
\,,\nonumber \\
b_{ij}&= -\pi T N_i\sum_{n=0}^{N_d}
\frac{\gamma_{ij} }{(\omega_n+\gamma_{ij}+\gamma_{ji})^4}
\\ +
\pi &T N_i\sum_{n=0}^{N_d}
\frac{\gamma_{ij}(\gamma_{ij}+\gamma_{ji})(\omega_n(\gamma_{ij}+\gamma_{ji})
+2\gamma_{ij}\gamma_{ji})}
{\omega_n^3(\omega_n+\gamma_{ij}+\gamma_{ji})^4}
\,,\nonumber
}
and 
\SubAlign{Eq:GLparameters:C}{
c_{ii}&=\pi T N_i \\ 
&\sum_{n=0}^{N_d}
\frac{  \gamma_{ij}(\omega_n+\gamma_{ji})
(\omega_n^2 +(\omega_n+\gamma_{ji})(\gamma_{ij}+\gamma_{ji})) }
{\omega_n^3(\omega_n+\gamma_{ij}+\gamma_{ji})^4}
\,,\nonumber \\
c_{ij}&=\pi T N_i\sum_{n=0}^{N_d}
\frac{  \gamma_{ij}(\omega_n+\gamma_{ji})(\omega_n+\gamma_{ij})
(\gamma_{ij}+\gamma_{ji}) }
{\omega_n^3(\omega_n+\gamma_{ij}+\gamma_{ji})^4}
\,.
}
Thus for a given set of input microscopic parameters, $\lambda_{ij}$, 
$\Gamma$ and $T$ close to $T_c$, we can reconstruct the coefficients 
\Eqref{Eq:GLparameters:K}--\Eqref{Eq:GLparameters:C} and investigate 
the ground-state properties of the GL theory by minimizing the free 
energy \Eqref{Eq:FreeEnergy} with respect to $|\Delta_j|$ and $\theta_{12}$.

\section{Ginzburg-Landau coefficients of the mixed-gradients-free basis}
\label{App:CoefficientsRewritten}

Rewriting the original Ginzburg-Landau model \Eqref{Eq:FreeEnergy}, using a 
linear combination of the components of the order parameter:
\SubAlign{Eq:NewFields:App}{
\psi_1&=\sqrt{k_{11}}\Delta_1+\sqrt{k_{22}}\Delta_2\,,  \\
\psi_2&=\sqrt{k_{11}}\Delta_1-\sqrt{k_{22}}\Delta_2\,,
}
allows a much simpler form of the kinetic terms which is convenient to 
investigate physical length scales. Within the new basis, the coefficients 
for the kinetic term of the rewritten Ginzburg-Landau functional  
\Eqref{Eq:FreeEnergy:Rotated} read as: 
\Equation{Eq:CoeffKinetic:App}{
\kappa_1=\frac{\sqrt{k_{11}k_{22}}+ k_{12}}{2\sqrt{k_{11}k_{22}}}
~~~~\text{and}~~~~
\kappa_2=\frac{\sqrt{k_{11}k_{22}}- k_{12}}{2\sqrt{k_{11}k_{22}}}
\,.
}
The coefficients of the potential read as
\SubAlign{Eq:Coeffients2:A}{ 
\alpha_{11}&=	\frac{a_{11}k_{22}+a_{22}k_{11}
	+2a_{12}\sqrt{k_{11}k_{22}}}{4k_{11}k_{22}}\,,	\\
\alpha_{22}&=	\frac{a_{11}k_{22}+a_{22}k_{11}
	-2a_{12}\sqrt{k_{11}k_{22}}}{4k_{11}k_{22}}\,,	\\
\alpha_{12}&=\frac{a_{11}k_{22}-a_{22}k_{11}}{4k_{11}k_{22}}							\,, 
}
and
\SubAlign{Eq:Coeffients2:B}{
\beta_{11}&=\frac{b_{11}k_{22}^2+b_{22}k_{11}^2}{16k_{11}^2k_{22}^2}
+\frac{b_{12}+c_{12}}{8k_{11}k_{22}}
+\frac{c_{11}k_{22}+c_{22}k_{11}}{4(k_{11}k_{22})^{3/2}}	\,,\\
\beta_{22}&=\frac{b_{11}k_{22}^2+b_{22}k_{11}^2}{16k_{11}^2k_{22}^2}
+\frac{b_{12}+c_{12}}{8k_{11}k_{22}}
-\frac{c_{11}k_{22}+c_{22}k_{11}}{4(k_{11}k_{22})^{3/2}}	\,,\\
\beta_{12}&=\frac{b_{11}k_{22}^2+b_{22}k_{11}^2}{8k_{11}^2k_{22}^2}	
-\frac{c_{12}}{4k_{11}k_{22}}		\,.~
}
Finally
\SubAlign{Eq:Coeffients2:C}{
\gamma_{11}&=\frac{b_{11}k_{22}^2-b_{22}k_{11}^2}{16k_{11}^2k_{22}^2}
+\frac{c_{11}k_{22}-c_{22}k_{11}}{8(k_{11}k_{22})^{3/2}}\,,		\\
\gamma_{22}&=\frac{b_{11}k_{22}^2-b_{22}k_{11}^2}{16k_{11}^2k_{22}^2}
-\frac{c_{11}k_{22}-c_{22}k_{11}}{8(k_{11}k_{22})^{3/2}}\,,		\\
\gamma_{12}&=\frac{b_{11}k_{22}^2+b_{22}k_{11}^2}{16k_{11}^2k_{22}^2} 
+\frac{c_{12}-b_{12}}{8k_{11}k_{22}} \,.
}

\clearpage
\section{Perturbations in the original basis}
\label{App:Eigenvectors}

Using the equations \Eqref{Eq:expansion} and \Eqref{Eq:expansion:original}, 
the perturbations are reconstructed, and they are expressed in terms of 
the perturbations in the \emph{rotated} basis, as
\SubAlign{Eq:expansion:original:def}{
\delta|\Delta_{1}|^2&=\frac{1}{2k_{11}}\left(\frac{f_1^2}{\kappa_1^2}
					+\frac{f_2^2}{\kappa_2^2}\right)   
				  \\
&-\left(\frac{u_1f_2}{\kappa_2}+\frac{u_2f_1}{\kappa_1}\right)
\sqrt{\frac{\kappa_1u_1^2+\kappa_2u_2^2}{k_{11}^2\kappa_1^3\kappa_2^3u_1^2u_2^2}}
\sin\bvarphi\phi		\nonumber\\
&-\frac{1}{k_{11}\kappa_1\kappa_2}\left(\left(f_1f_2+
\frac{\kappa_1u_1^2+\kappa_2u_2^2}{2u_1u_2} \right)\cos\bvarphi
\right)  \,,\nonumber  \\
\delta|\Delta_{2}|^2&=\frac{1}{2k_{22}}\left(\frac{f_1^2}{\kappa_1^2}
					+\frac{f_2^2}{\kappa_2^2}\right)     \\
&+\left(\frac{u_1f_2}{\kappa_2}+\frac{u_2f_1}{\kappa_1}\right)
\sqrt{\frac{\kappa_1u_1^2+\kappa_2u_2^2}{k_{22}^2\kappa_1^3\kappa_2^3u_1^2u_2^2}}
\sin\bvarphi\phi		\nonumber\\
&+\frac{1}{k_{22}\kappa_1\kappa_2}\left(\left(f_1f_2+
\frac{\kappa_1u_1^2+\kappa_2u_2^2}{2u_1u_2} \right)\cos\bvarphi
\right)  \,,\nonumber  \\
\delta\theta_{12}&=
\frac{-2u_1u_2(u_1^2+u_2^2)\cos\bvarphi\sqrt{\frac{\kappa_1u_1^2+\kappa_2u_2^2}
{\kappa_1\kappa_2u_1^2u_2^2}}\phi}{(u_1^2+u_2^2)^2+4u_1^2u_2^2\sin^2\bvarphi}
\nonumber\\
&~-\frac{2u_1u_2(u_1-u_2)\left(\frac{u_1f_2}{\kappa_2}-\frac{u_2f_1}{\kappa_1}\right)}
{(u_1^2+u_2^2)^2+4u_1^2u_2^2\sin^2\bvarphi}		\,.
}
\section{Second critical field}
 \label{App:Hc2}

$\Hc{2}$ can be obtained by considering the perturbation operator around 
the normal state. More precisely, in the original parametrization the 
normal state is $|\Delta_1|=|\Delta_2|=0$. Using \Eqref{Eq:Transform}, 
this implies that the normal state in the new variables is 
$|\psi_1|=|\psi_2|=0$ and thus $u_1=u_2=0$ and $\bvarphi=0$.

\begin{figure}[!htb]
\hbox to \linewidth{ \hss
\includegraphics[width=0.99\linewidth]{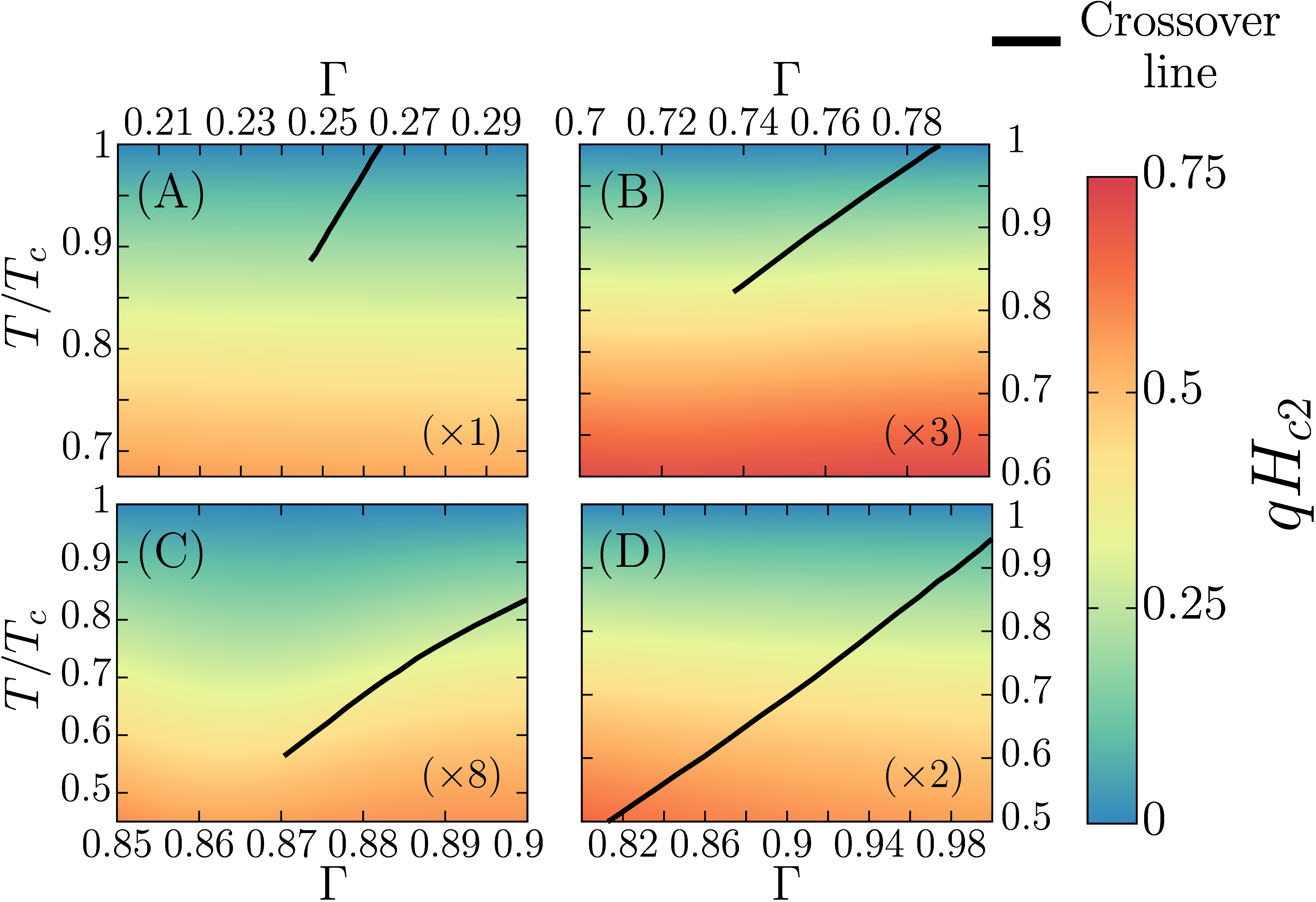}
\hss}
\caption{
Upper critical field $\Hc{2}$, as defined in \Eqref{Eq:hc2}, as a function 
of temperature and interband scattering $\Gamma$, for equal diffusivities 
in both bands ($r_d=1$). 
The different panels correspond to different values of the coupling matrix 
$\hat{\lambda}$ that are displayed in \Figref{Fig:Diagram}. Clearly, the 
complexity of the phase diagram of the model has little effect on the behavior 
of $\Hc{2}$.
}
\label{Fig:Hc2}
\end{figure}

Close to the second critical field $\Hc{2}$ the magnetic field is 
approximately constant: $\B=B_0{\bs e}_z$ and the densities are small. 
Thus the Ginzburg-Landau equations \Eqref{Eq:EOM:Rotated} can be linearized 
around the normal state as 
\Equation{Eq:linearized:hc2:1}{
\Pi^2\Upsilon={\cal M}^2\big\rvert_{u_1=u_2=\bvarphi=0}\Upsilon
\equiv{\cal M}_0^2\Upsilon\,.
}
Because of the gauge invariance, the vector potential can be parametrized 
in the Landau Gauge as $\A=(0,B_0x,0)^{-1}$. As a result, the linearized 
Ginzburg-Landau equations read as 
\Equation{Eq:linearized:hc2:2}{
\left(\Grad^2-\left(qB_0x\right)^2\right)\Upsilon
={\cal M}_0^2\Upsilon\,.
}
For the simple Gaussian ansatz $\Upsilon=C\exp\left(-\frac{x^2}{2\xi^2}\right)$ 
with the vector $C=(C_1,C_2)^T$ and $q B_0=1/\xi^2$. The equation 
\Eqref{Eq:linearized:hc2:2} further simplifies:
\Equation{Eq:linearized:hc2:3}{
{\cal M}_0^2\Upsilon=\frac{-1}{\xi^2}\Upsilon\,.
}
Thus $1/\xi^2$ is an eigenvalue of $-{\cal M}_0^2$. More precisely, its 
largest: 
\Equation{Eq:linearized:hc2:4}{
q\Hc{2}=\frac{1}{\xi^2}:=
\mathrm{max}\Big(\mathrm{Eigenvalue}\big[-{\cal M}_0^2\big] \Big)\,.
}

It is easy to realize that the perturbations of the relative phase $\Upsilon$ 
decouple from density perturbations. The mass matrix thus becomes:
\Equation{Eq:perturbation:normal}{
{\cal M}_0^2=2\left(\begin{array}{cc}
\alpha_{11}/\kappa_1 			& \alpha_{12}/\sqrt{\kappa_1\kappa_2}	\\ 
\alpha_{12}/\sqrt{\kappa_1\kappa_2}	& \alpha_{22}/\kappa_2 	
\end{array}\right)\,,
}
and its eigenvalues are 
\Equation{Eq:perturbation:normal:EV}{
\frac{\kappa_2\alpha_{11}+\kappa_1\alpha_{22}
\pm\sqrt{(\kappa_2\alpha_{11}-\kappa_1\alpha_{22})^2
		+4\alpha_{12}^2\kappa_1\kappa_2} }{\kappa_1\kappa_2}\,.
}
As a result, we find the second critical field in the dimensionless units 
of Eq.~\Eqref{Eq:FreeEnergy} 
\Equation{Eq:hc2}{
q\Hc{2}=-\frac{\alpha_{11}}{\kappa_1}
			-\frac{\alpha_{22}}{\kappa_2}
+\sqrt{\left(\frac{\alpha_{11}}{\kappa_1}-\frac{\alpha_{22}}{\kappa_2}\right)^2+4\frac{\alpha_{12}^2}{\kappa_1\kappa_2} }\,.
}

\Figref{Fig:Hc2} shows the upper critical field $\Hc{2}$, defined in 
\Eqref{Eq:hc2}, as a function of temperature and interband impurity 
scattering $\Gamma$, in the case of equal diffusivities. This shows 
the case of two-band superconductors with different intraband and 
interband coupling.

\clearpage
\section{Effect of the relative diffusion constant}
\label{App:rd}

\begin{figure}[!tb]
\hbox to \linewidth{ \hss
\includegraphics[width=\linewidth]{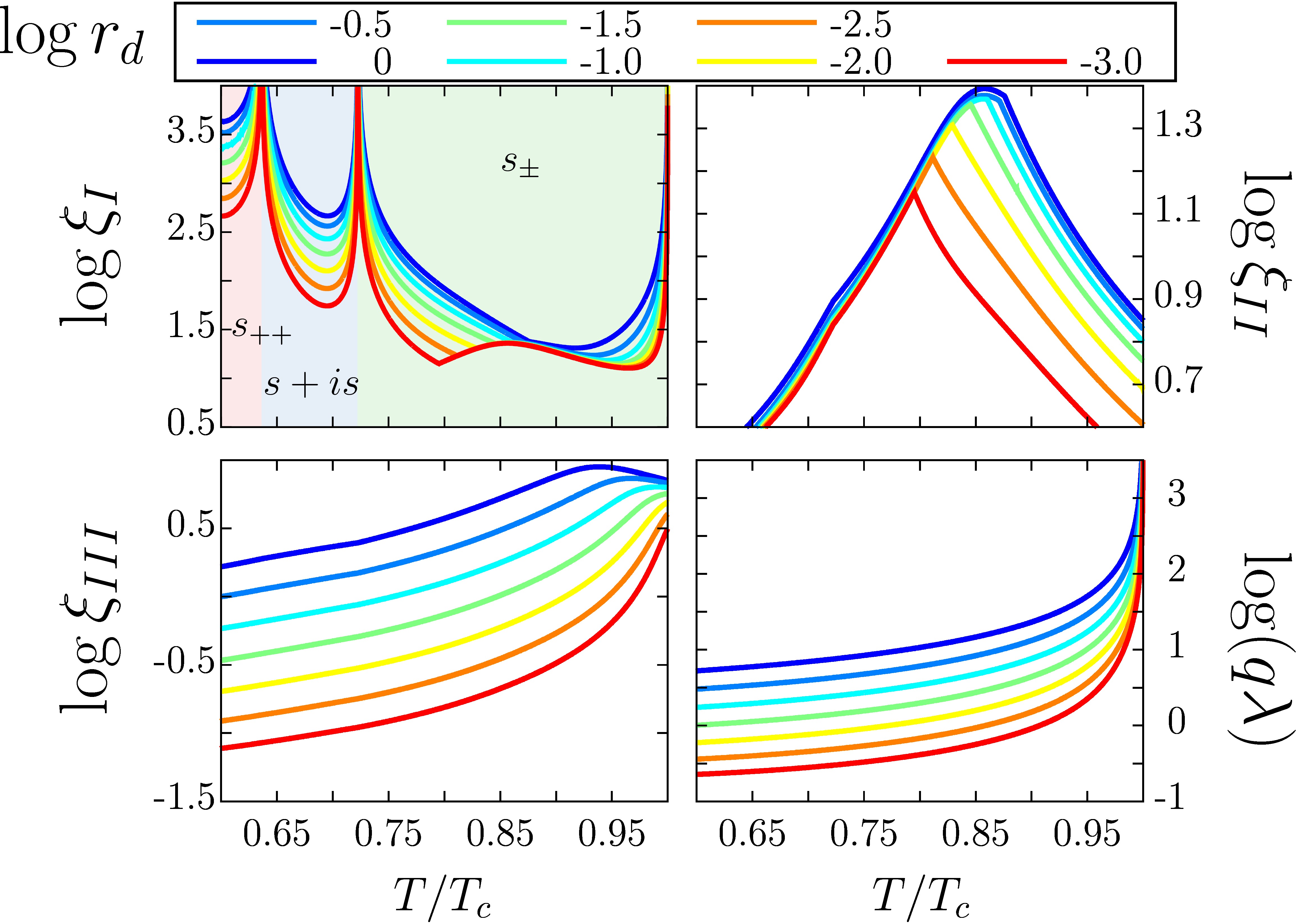}
\hss} \vspace{0.1cm}
\hbox to \linewidth{ \hss
\hspace{-.35cm}
\includegraphics[width=0.8125\linewidth]{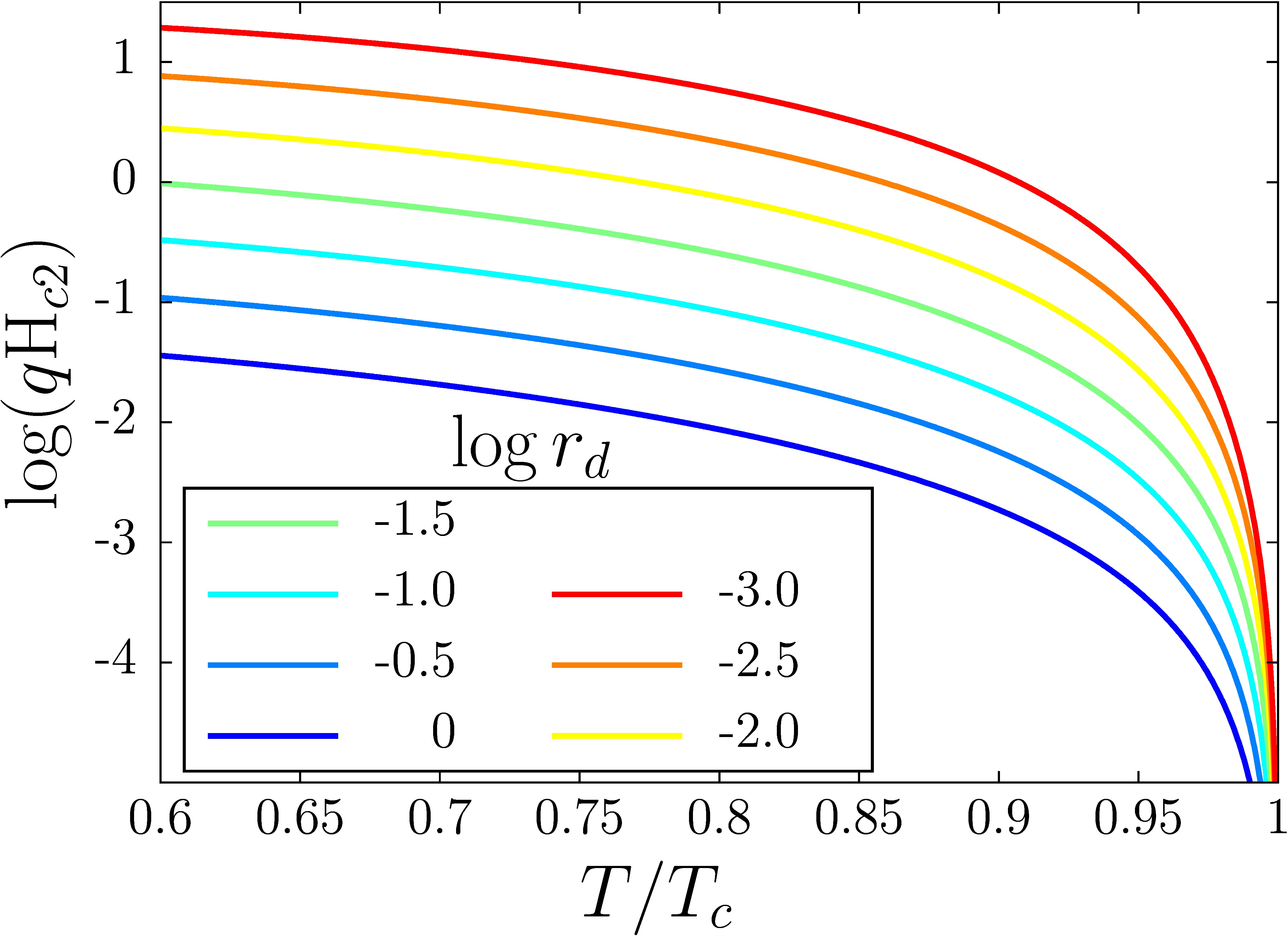}
\hss}
\caption{
Effect of relative diffusion constant on the different length scales
and on the upper critical field $\Hc{2}$ (corresponding to panel (B) of 
diagram \Figref{Fig:Diagram}). The top block displays the length scales 
as functions of the temperature for a given interband scattering $\Gamma=0.7275$
The relative diffusion constant 
affects only quantitatively the different length 
scales and do not affect the phase diagram. 
Similarly, the bottommost panel shows upper critical field $\Hc{2}$ as 
a function of the temperature for the same parameter set. Increasing 
the relative diffusion constant
also increases the upper critical field.
}
\label{Fig:Diffusivity-ratio}
\end{figure}

Figure~\ref{Fig:Diffusivity-ratio} shows the effect of 
the relative diffusion constant
on the different length scales and on the upper critical 
field $\Hc{2}$, for a dirty two-band superconductor with nearly degenerate 
bands and intermediate repulsive interband pairing interaction (corresponding 
to panel (B) of diagram \Figref{Fig:Diagram}) . 
Relative diffusion constant
has only a quantitative influence on the various length scales, while it 
increases the upper critical field $\Hc{2}$.

%

\end{document}